\documentclass[%
reprint,
nofootinbib,
amsmath,amssymb,
aps,
]{revtex4-2}
\usepackage[colorlinks=true,linkcolor=blue,citecolor=blue,urlcolor=
blue]{hyperref}

\usepackage{graphicx}
\usepackage{dcolumn}
\usepackage{bm}
\usepackage{amsmath}
\usepackage[dvipsnames]{xcolor}
\usepackage{subcaption}
\usepackage{lipsum}
\usepackage{mwe}
\usepackage{multirow}

\begin{document}
	
	\preprint{APS/123-QED}
	\title{Investigating the Role of Electric Fields on Flow Harmonics in Heavy-Ion Collisions}
	
	\author{Ankit Kumar Panda}
	\email{ankitkumar.panda@niser.ac.in}
	\author{Reghukrishnan Gangadharan}
	\email{reghukrishnang@niser.ac.in}%
	\author{Victor Roy}%
	\email{victor@niser.ac.in}
	\affiliation{School of Physical Sciences, National Institute of Science Education and Research,
		An OCC of Homi Bhabha National Institute, Jatni-752050, India}%

	\date{\today}
	
	\begin{abstract}
Using the blast-wave model, we explore the effect of electric fields on spectra and flow harmonics (especially the elliptic flow) for charged pions and protons. We incorporate the first-order correction to the single-particle distribution function due to the electric fields and the dissipative effect while calculating the invariant yields of hadron in the Cooper-Frey prescription at the freezeout hypersurface. We find a noticeable correction to the directed and elliptic flow of pions and protons for unidirectional and azimuthal asymmetric electric fields in the transverse plane of magnitude $\sim m_{\pi}^{2}$. Further, we observe mass dependency of the directed flow generated due to the electric fields. The splitting of particle and antiparticle's elliptic flow is also discussed.	
	\end{abstract}
	
	\maketitle
	\section{Introduction} 
	In high energy heavy-ion collisions at RHIC and LHC, two highly Lorentz contracted nuclei colliding onto one another produces a hot and dense novel state of matter known as Quark Gluon Plasma~(QGP). The QGP formed in heavy-ion collisions is the same as the medium that existed during the microsecond old universe, which consisted of almost freely moving quarks and gluons that are usually confined in colorless hadrons at low or zero temperatures~\cite{PhysRevD.10.2445, Karsch_2000}. Hydrodynamic calculations have confirmed that the QGP is a strongly coupled fluid with extremely low $\frac{\eta}{s}$ value {\cite{PhysRevLett.99.172301,PhysRevC.77.064901,PhysRevLett.94.111601}}. Along with the QGP, an intense electromagnetic field of the order of $10^{14}$ T is generated in the initial stage of heavy ion collision in Au+Au collision at top RHIC energies {\cite{Kharzeev:2007jp,Skokov:2009qp,PhysRevC.83.039903,PhysRevC.83.054911,PhysRevC.85.044907,Tuchin:2013ie,McLerran:2013hla}}. 
The primary source of intense electromagnetic (EM) fields in heavy-ion collisions are the spectator nucleons (the nucleons that are unaffected by collisions). 

Relativistic viscous hydrodynamics is a well-established formalism for describing the space-time evolution of the quark-gluon plasma (QGP). However, owing to the finite electrical conductivity of the high temperature QGP and low temperature hadronic gas phase both phases evolving in intense EM fields, the proper framework, we believe, is the relativistic resistive magnetohydrodynamics (RRMHD).
The straightforward generalization of non-relativistic viscous hydrodynamics to the relativistic regime leads to an acausal theory.
The acausality problem is usually cured by incorporating higher-order gradient terms in the entropy-four current or in the energy-momentum tensor of the fluid~\cite{ISRAEL1979341}.  The space-time evolution of the QGP should be studied using a causal RRMHD theory. There have been recent developments on the formulation of causal (second-order) magnetohydrodynamic formalisms from the kinetic theory perspective {\cite{Denicol:2018rbw,Denicol:2019iyh,Panda:2020zhr,Panda:2021pvq,Panda:2022nsw,Most:2021rhr}}. 
Most of these studies have reported a few new transport coefficients arising due to the external EM fields, some of which were non-dissipative as they did not increase entropy. However, a systematic investigation of many of these transport coefficients has not yet been done.
Moreover, the effect of electric fields on bulk observables such as transverse momentum ($p_T$) spectra and flow coefficients of charged hadrons has not been extensively studied, to our knowledge, except for a few studies~\cite{PhysRevC.89.054905,PhysRevC.98.055201}. Developing a RRMHD numerical code is a challenging task~\cite{Nakamura:2022wqr,Inghirami:2016iru}, and it is known that viscosity (and EM fields) affects the space-time evolution of the fluid and also the one-particle thermal distribution of particles used in the Cooper-Frye formulation, which ultimately gives the final experimentally observed particles from the fluid elements. 

Here, we use a simple but effective blastwave model~\cite{Schnedermann:1993ws} to investigate the influence of the electric field on experimental bulk observables such as first and second-order flow harmonics ($v_1, v_2$) and transverse momentum ($p_T$) spectra of charged hadrons, due to corrections only in the Cooper-Frye prescription. These observables are known as `bulk observables' because they involve the collective motion of many particles in the fluid.The effect of viscous corrections is introduced through a relaxation time $\tau_c$, which is a free parameter in our case. We note that a more accurate estimate of the effect of EM fields on bulk observables requires a magnetohydrodynamic evolution {\cite{Nakamura:2022wqr}} of the fluid.
It is important to note that in addition to second-order flow harmonics ($v_2$) and transverse momentum ($p_T$) spectra of charged hadrons, the electromagnetic field may result in a variety of other novel phenomena such as the Chiral Magnetic Effect and Chiral Separation Effect~\cite{PhysRevD.78.074033, PhysRevLett.107.052303,PhysRevD.83.085007,PhysRevLett.107.052303}. Other works that study the effect of EM fields on the QGP include~\cite{Zhang:2022lje,Inghirami:2019mkc, Roy:2017yvg,Gezhagn:2021oav, Huang:2015oca}.

The article is organized as follows: we discuss the blast wave model in Section~\ref{blastwave}, then we discuss the Cooper-Frye mechanism and other relevant formulas, along with the setup, in Sections~\ref{cooperfreyformalism} and~\ref{setup}. Results are discussed in Section~\ref{results}. Finally, we conclude and summarize our study in Section~\ref{conclusion}. Throughout the article, we use natural units where $\hbar=c=k_\mathrm{B}=1$. The metric signature used here is $g_{\mu\nu}=\mathrm{diag}(+,-,-,-)$.			
	
	\section{Blast wave model}{\label{blastwave}}
The blast-wave model is a simple model that considers the collective motion of the matter produced in heavy-ion collisions and parameterizes its four-velocity. It further assumes that hadrons are produced from a constant temperature freeze-out hypersurface, with the freeze-out temperature being a free parameter. The invariant yields of hadrons are obtained from the Cooper-Frye formalism, which is described later. Despite its simplicity, the blast-wave model can successfully describe experimental data qualitatively and quantitatively in most cases.		

In heavy-ion collisions, one popular parametrization of the fluid four-velocity is inspired by the Bjorken model of boost-invariant expansion in the longitudinal direction of the fluid. In this work, we use the Milne coordinate $\left(\tau,\eta,r,\phi\right)$ with the metric $g_{\mu\nu}=\mathrm{diag}\left(1,-\tau^2,-1,-r^2\right)$, and the transformation between the Cartesian and Milne coordinate is given as	
	\begin{gather}\nonumber
		\begin{bmatrix} t  \\ x \\ y \\ z \end{bmatrix}
		=
		\begin{bmatrix}
			cosh \eta & 0 & 0 & 0 \\
			0 & 0 & cos \phi & 0 \\
			0 & 0 & sin \phi & 0 \\
			sinh \eta & 0 & 0 & 0
		\end{bmatrix}
		\begin{bmatrix} \tau  \\ \eta \\ r \\ \phi \end{bmatrix}.
	\end{gather}
Where	
	\begin{eqnarray}\nonumber
		\tau&=&\sqrt{t^2-z^2},\\ \nonumber
		\eta&=&\tanh^{-1}{z/t},\\ \nonumber
		r&=&\sqrt{x^2+y^2},\\ \nonumber
		\phi&=&arctan2(y,x).
	\end{eqnarray}
	
	The parameterized form of the velocity four vector with longitudinal boost-invariance is given as
	\begin{eqnarray} \nonumber
		u^{r}&=&u_0 \frac{r}{R}\left[1+2\sum_{n=1}^{\infty} c_n \cos{n\left[\phi-\psi_n\right]}\right]\Theta\left(R-r\right),\\ \nonumber
		u^{\phi}&=&u^{\eta}=0 , \\ \label{parametrisedvelocity}
		u^{\tau}&=&\sqrt{1+\left(u^r\right)^2} .
	\end{eqnarray} 
	
	Where, $u^{\tau},u^{r}, u^{\phi},u^{\eta}$ are the components of the fluid four velocity;  $u_0$, and $c_n$'s are free parameters used to reproduce the $p_{T}$ spectra (invariant yield) and the flow harmonics of charged hadrons. $\Theta\left(R-r\right)$ is the Heaviside function which imposes the condition that if $r>R$, $u^{r}=0$. R is the radius of the freezeout hypersurface, $\phi$ is the azimuthal angle in co-ordinate space and $\psi_n$ is the $n \mbox{-} th$ order participant plane angle. As we do not consider event-by-event fluctuations in this work, we set $\psi_n=0$ i.e., the minor axis of the participant planes coincide with the direction of the impact parameter. We parametrised the temperature on the freezeout hypersurface as
	\begin{eqnarray}\nonumber
		T\left(\tau,\eta,r,\phi\right)&=&T_0 \Theta\left(R-r\right).
	\end{eqnarray}
	Where $T_0$  denotes the temperature at the freezeout hypersurface.

	\section{Cooper-frey formalism} \label{cooperfreyformalism}
As mentioned earlier, the invariant yield of hadrons is obtained from the Cooper-Frye formula~\cite{PhysRevD.10.186}, which assumes that the freeze-out hypersurface is a timelike vector $d\Sigma_{\mu}$ given by $\left(\tau d\eta dr rd\phi,0,0,0\right)$. The invariant yield is given by the following equation:
\begin{eqnarray} \label{eq:invYield}
\frac{dN}{d^2p_{T} dy}&=&\frac{\mathcal{G}}{\left(2\pi\right)^3}\int p^{\mu}d\Sigma_{\mu}f(x,p).
\end{eqnarray}
Here, $f(x,p)$ is the single-particle distribution function, and $x$ and $p$ are the position and momentum four-vectors of the particles, respectively. $\mathcal{G}$ is the degeneracy factor.

If the system is not in local thermal equilibrium, as is the case for a rapidly expanding fireball, the single-particle distribution function must take into account the deviation from equilibrium when calculating hadron yields using the Cooper-Frye prescription. The distribution function is usually decomposed into an equilibrium part $f_{0}$ and a small non-equilibrium part $\delta f$, so that the total distribution function becomes $f=f_{0}+\delta f$. The second-order correction to $f$ gives rise to some new transport coefficients due to the external magnetic field. The temperature and mass dependence of some of these transport coefficients were discussed in~\cite{Das:2022lqh}, and we will show the results for the remaining coefficients later in the Results section.

However, in this work, we consider terms up to first-order in gradients~\cite{Panda:2021pvq}. In this case, the invariant yield (Eq.~\eqref{eq:invYield}) becomes

\begin{eqnarray}\label{cooperfrey}
		\frac{dN}{d^2p_{T} dy}&=&\frac{\mathcal{G}}{\left(2\pi\right)^3}\int p^{\mu}d\Sigma_{\mu} \left(f_0 + \delta f ^{1}\right).
\end{eqnarray}
Where $\delta f^{1} \ll$  $f_0$. In \cite{Panda:2020zhr,Panda:2021pvq} $ \delta f^{1}$ is calculated using the Boltzmann equation using the Relaxation Time Approximation~(RTA). We give the expression for $ \delta f^{1}$ in Appendix.\eqref{deltaf} for the sake of completeness.

	\section{Setup}\label{setup}
\begin{center}
		\begin{figure}[ht] 
			\begin{center}
				\includegraphics[width=0.5\textwidth,,height=0.25\textheight]{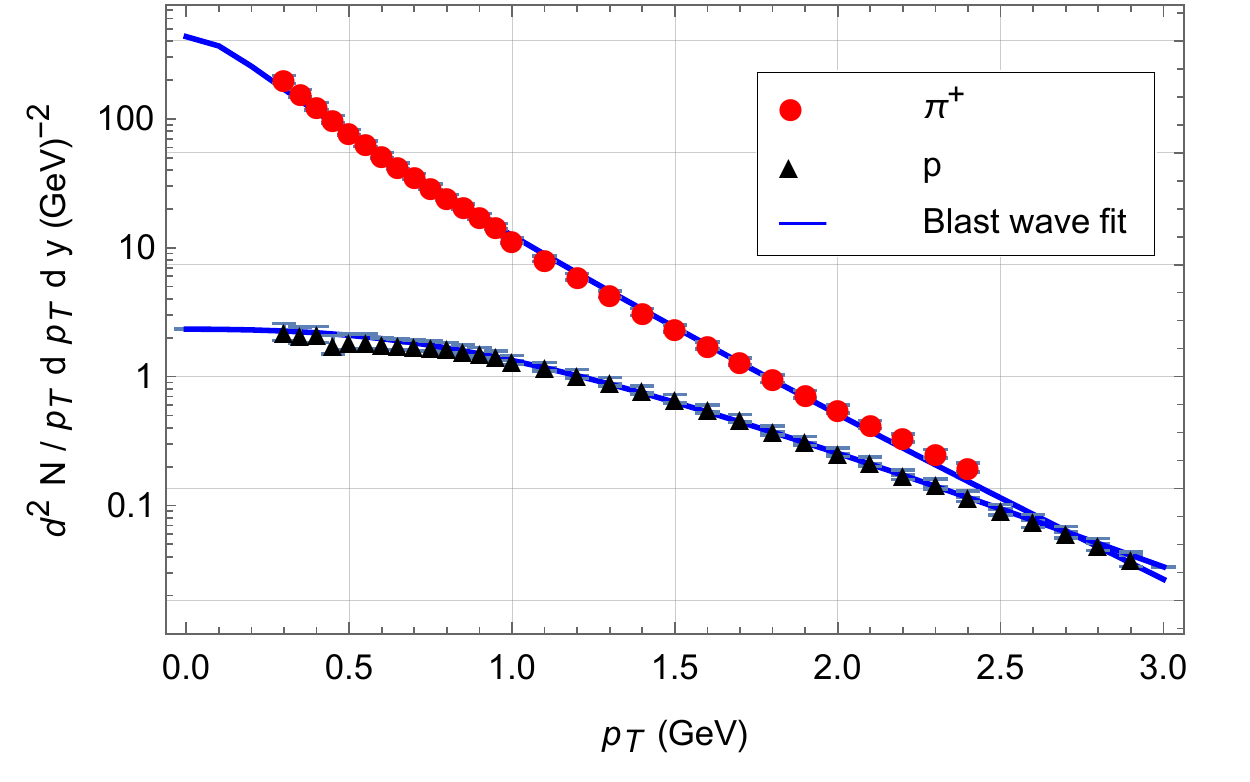} 
				\\ \hfill
				\includegraphics[width=0.5\textwidth,,height=0.25\textheight]{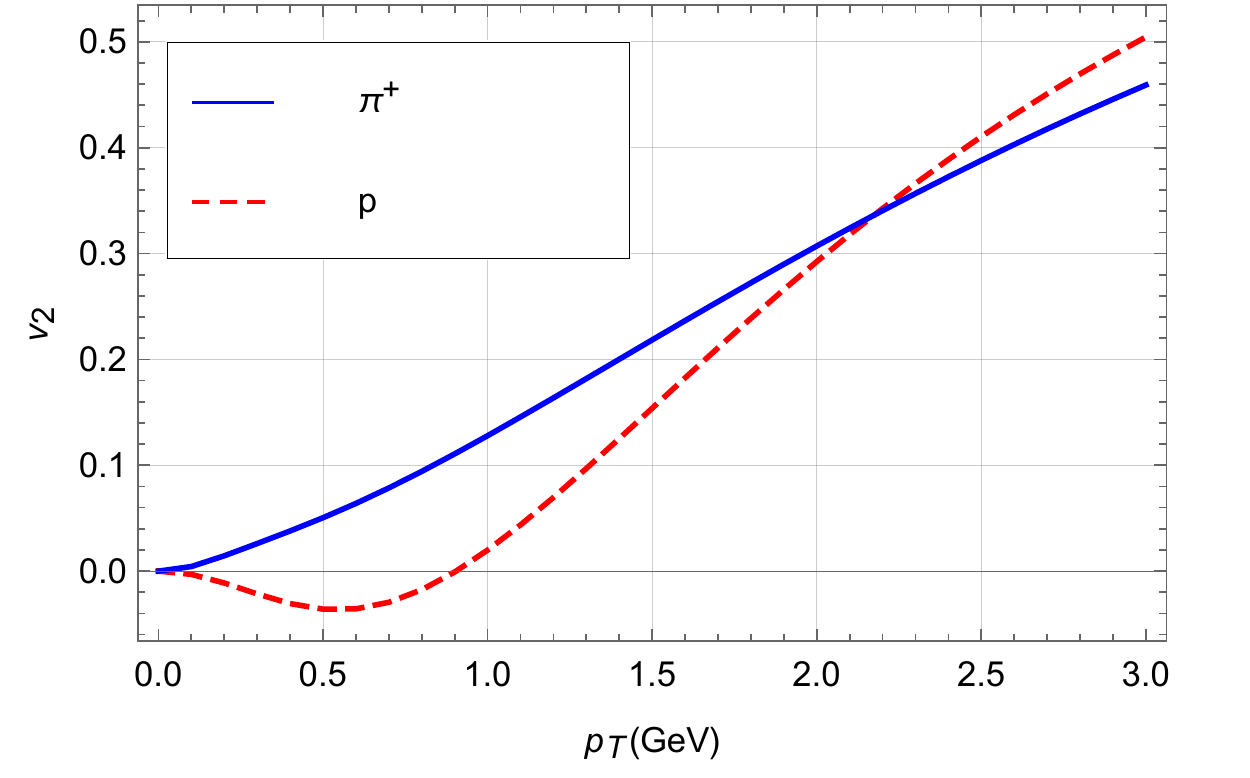} 
				\caption{(Color online)Top plot is the comparison of the experimentally measured $\pi^{+}$ (red filled circles) and protons (black triangles) for Pb+ Pb collision at $\sqrt{s}$=2.76 TeV for 20-30 $\%$ centrality  with the ideal blast wave results (lines) for the parameters in Table~\eqref{table:paramset}.The lower panel shows the  corresponding $v_2$ vs $p_{T}$ for $\pi^{+}$ (blue line) and  protons (red dashed line).} \label{fitplot}
			\end{center}
		\end{figure}
	\end{center}

	For the current study, we consider only $n=2$ in the expression for $u^{r}$ (Eq.\eqref{parametrisedvelocity}), resulting in \[u^{r}= u_0 \frac{r}{R}[1+2 c_2 \cos(2\phi)]\Theta\left(R-r\right).\] The equilibrium distribution is $f_{0}=\left(e^{\beta u\cdot p-\alpha}+r\right)^{-1}$, here $\beta$ is the inverse temperature,
	$r= \pm 1$ corresponds to  fermions and bosons respectively. 
We use the parameters given in Table~\eqref{table:paramset} to obtain the invariant yield of $\pi^{+}$ that matches the ALICE measurement shown as red circles in the top panel of Fig.(\ref{fitplot})\cite{ALICE:2013mez}. Earlier hydrodynamic model studies~\cite{Roy:2010qg} have shown that it is not possible to simultaneously describe $\pi^{+}$ and $\rm{p}$ spectra for zero baryon chemical potential, so we use a different set of values for $u_{0}$, $c_{2}$, and $T$ (given in the rightmost column of Table~\eqref{table:paramset}) to explain the proton spectra. The blast-wave model with these parameters reasonably well explains the experimental data. The bottom panel of Fig.~(\ref{fitplot}) shows the blast-wave results for $v_2$ of $\pi^{+}$ and $\rm{p}$. We do not compare these ideal results with experimental data since they are known to over-predict the experimental measurements, but we show them for the sake of qualitative description.	

\begin{figure}[h!]
		\includegraphics[width=0.5\textwidth]{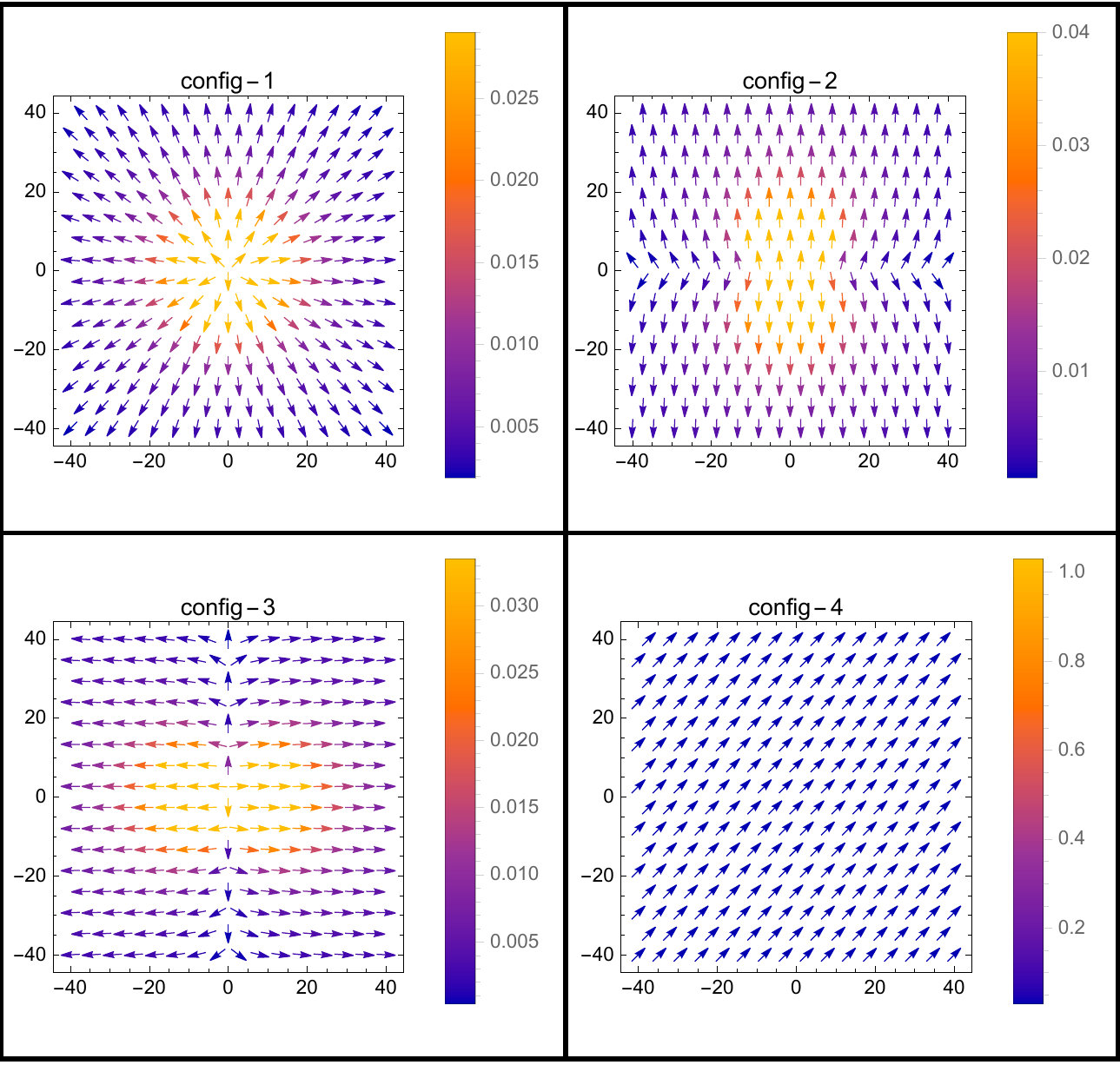}
		\\[\smallskipamount] \hfill
		\caption{(Color online) Electric field configurations in the transverse plane. Detail expressions for different configurations are given in Appendix.\eqref{electricfieldtransform}. Magnitude of electric fields (in $\rm{GeV}^{2}$) are shown using colour map. } {\label{electricfieldconfig}}
	\end{figure}

After setting up the parameters for the ideal case (zero viscosity), we can now explore the effect of viscosity (through the relaxation time $\tau_c$) and electric fields on spectra and flow harmonics by comparing the corresponding results with the ideal results. In our model, $\tau_c$ and the electric field $qE$ appear in $\delta f^{1}$, giving rise to additional corrections to the invariant yields and flow harmonics of charged hadrons. We compute the $p_T$ differential $n$-th order flow coefficient using the usual formula:		
	\begin{eqnarray}
		v_n (p_T,y) &=& \frac{\int^{2\pi}_{0} d\varphi \cos({n \varphi}) \frac{dN}{d^2p_{T} dy} }{\int^{2\pi}_{0} d \varphi \frac{dN}{d^2p_{T} dy}  } .
	\end{eqnarray}
	\begin{table}[ht]
		\begin{tabular}{|p{0.5cm}|p{1.8cm}|p{1.8cm}|}
			\hline
			&    $\pi^{+}$& $ {\rm{p}}$   \\ 
			\hline
			$u_0$ & 1.2 & 1.22\\
			\hline
			T & 130 MeV & 140 MeV\\
			\hline
			R& 10 fm & 10 fm \\
			\hline
			$ \tau $& 6 fm & 6 fm \\
			\hline
			m & 139.5 MeV & 938 MeV \\
			\hline
			$c_{2}$ & 0.1 & 0.15\\
			\hline
		\end{tabular}
		\\
		\caption{The fit parameters for $\pi^+$ and ${\rm{p}}$ respectively at mid-rapidity. }
		\label{table:paramset}
	\end{table}
	As we need the electric field distribution on the freezeout hypersurface to be used in the 
	Cooper-Frye formula (Eq.\eqref{eq:invYield}) we use a parameterized form of the EM field. In principle,
	the fields generated in the initial stages of heavy-ion collisions due to the charged protons inside the two colliding nucleus would 
	evolve with the QGP fluid, but the blast wave model does not allow any such self-consistent dynamical evolution of fields. The magnitude of the electric fields in all the cases are kept 
	Here we use parameterised electric fields of four different configurations in the transverse plane (XY plane) while calculating the invariant yields. Some of these configurations does not represent the actual scenario 
	encountered in heavy-ion collisions but we use them for exploratory purpose. 
	In the left panel of Fig.(\ref{electricfieldconfig}) we show the first setup of electric fields which we call {\bf{config-1}} henceforth, this represents isotropic fields due to a point charge of large magnitude at the origin, the right panel shows {\bf{config-2}} that closely resembles the field configuration expected in symmetric heavy-ion collision.  {\bf{config-3}} as shown in the left panel of Fig.(\ref{electricfieldconfig}) is the $\pi/2$ rotated version of {\bf{config-2}}. 
	The right panel of Fig.(\ref{electricfieldconfig}) represents a constant unidirectional electric fields which might be applied in a limited sense for large-small nuclei collisions, this is {\bf{config-4}}. To better understand the effect of electric fields on flow harmonics we note the following
	\begin{itemize}
	\item {\bf{config-1}}: isotropic fields; expected to unalter the flow harmonics,
	\item {\bf{config-2}}: prolate-like fields; expected to decrease the flow harmonics,
	\item {\bf{config-3}}: oblate-like fields; expected to increase the flow harmonics,
	\item {\bf{config-4}}: directional fields; expected to alter directional flow only. 
	\end{itemize}
	
	\section{Results}{\label{results}}
	Here we discuss our main results for different configurations of electric fields (described in the previous section) on the spectra ,and $v_2$  for pions and  protons.
	\begin{figure}[h!]
		\includegraphics[width=0.5\textwidth]{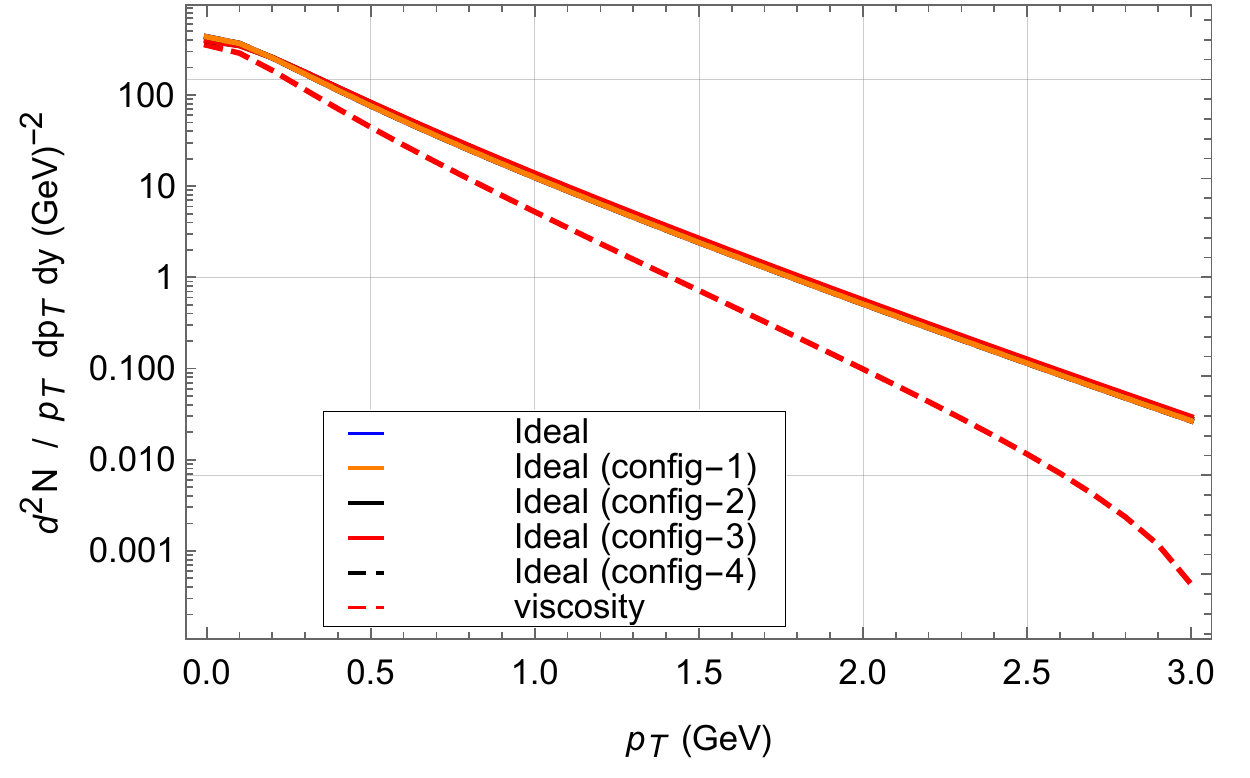} \\
		\includegraphics[width=0.5\textwidth]{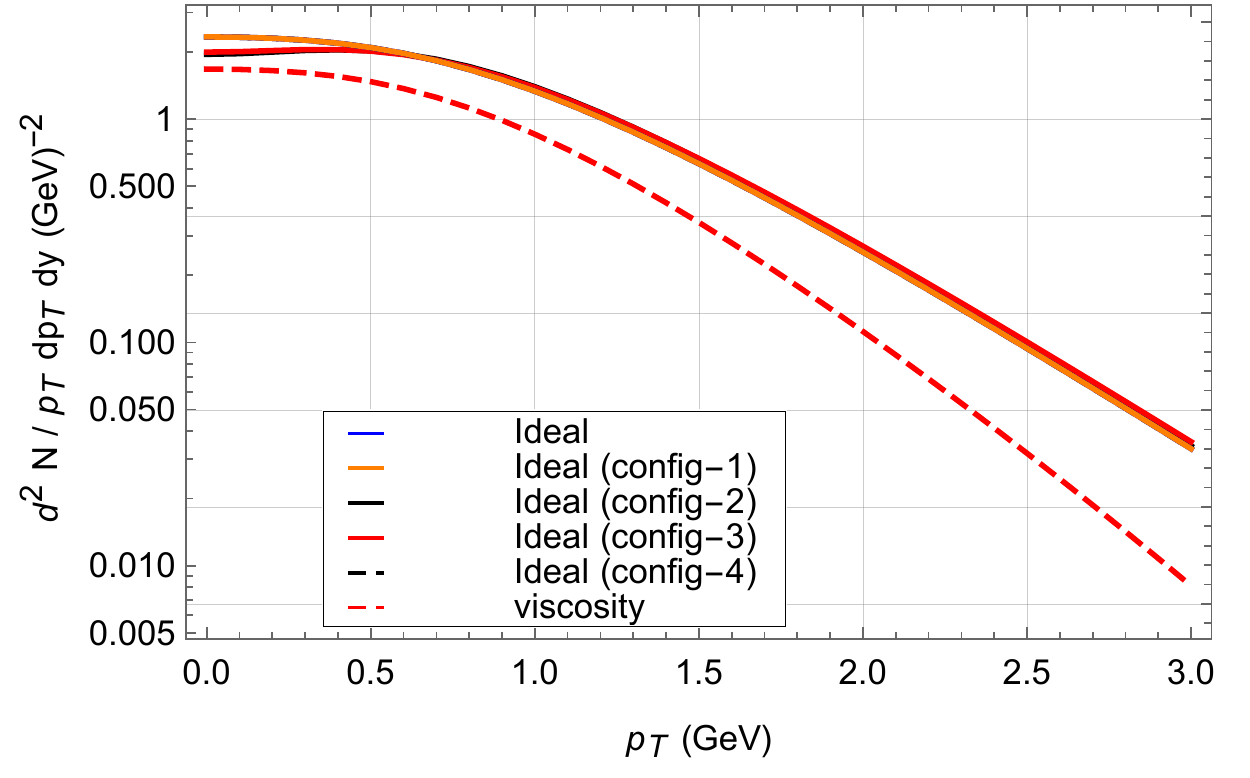}\\
		\caption{(Color online) Upper panel represents the $p_{T}$ Spectra for $\pi^{+}$ and lower panel for proton with different transverse field configurations as in Fig.\eqref{electricfieldconfig} and for the case of viscosity for $t_c = $ 1fm .} {\label{pionprotonspectra}}
	\end{figure}
	Top panel of Fig.(\ref{pionprotonspectra}) shows the dependence of $p_T$ spectra of $\pi^{+}$ on the electric fields for ideal case, for comparison we also show the results for non-zero viscosity but with no electric fields. The magnitude of viscosity in our model is controlled through $t_c$ which is set to 1 fm for the results shown here. Here we see that the $p_T$ spectra hardly shows any dependence on different transverse electric field configurations (solid orange, black, red lines correspond to {\bf{config-1}}, {\bf{config-2}}, and {\bf{config-3}} respectively). As expected the effect of finite viscosity is comparatively more prominent on the $p_T$ spectra; we see a suppression at higher $p_T$ region for the viscous case. Since both bulk and shear viscosity is present 
in our case, the slope of the spectra is determined by the relative contributions of these two viscosities~\cite{PhysRevC.90.034907,Denicol:2010tr}.The bottom panel of Fig.(\ref{pionprotonspectra}) shows the $p_{T}$ spectra for protons for ideal (with electric fields) and viscous (without electric field) cases where we see a small suppression in the lower $p_T$ region for the case of {\bf{config-2}} and {\bf{config-3}} for ideal case. We also see a suppression at higher $p_T$ for viscous case as was seen for $\pi^{+}$.
	\begin{figure}[h!]
		\includegraphics[width=0.45\textwidth]{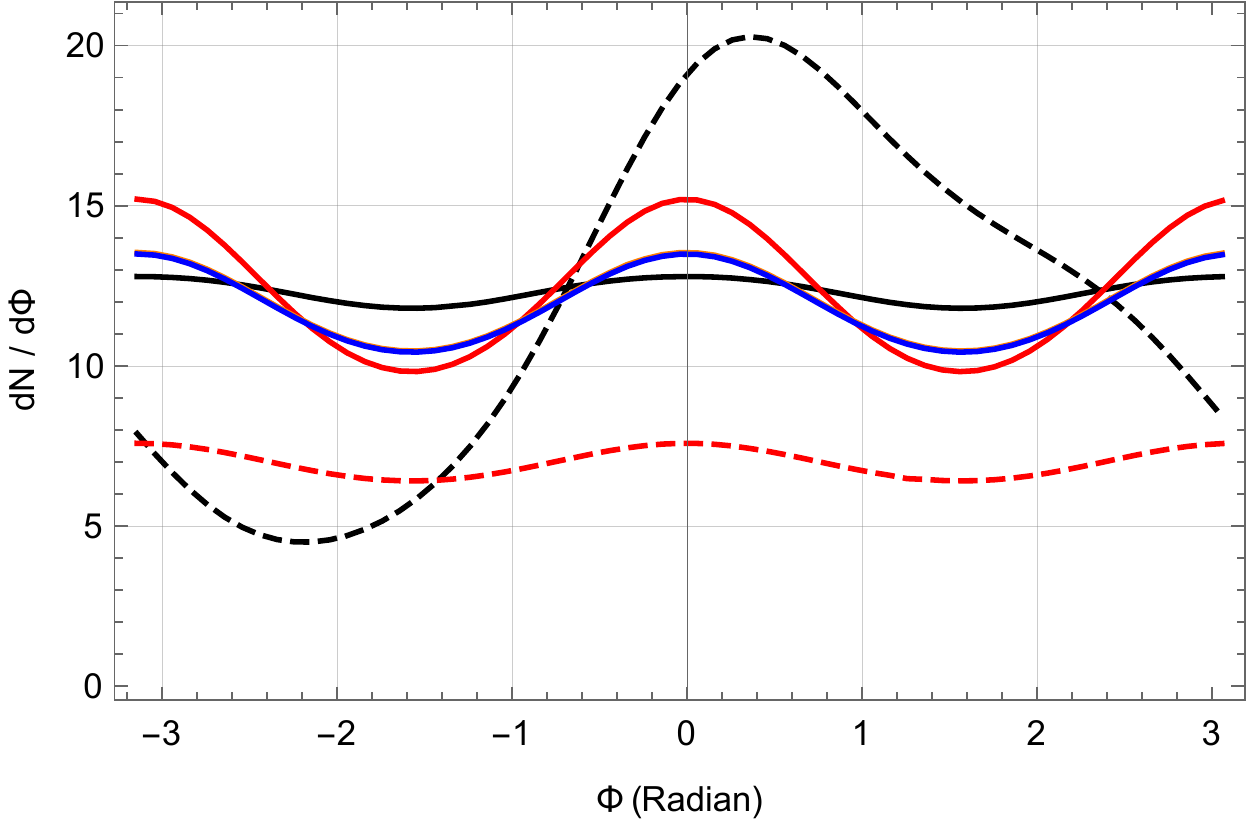} \\ 
		\includegraphics[width=0.5\textwidth]{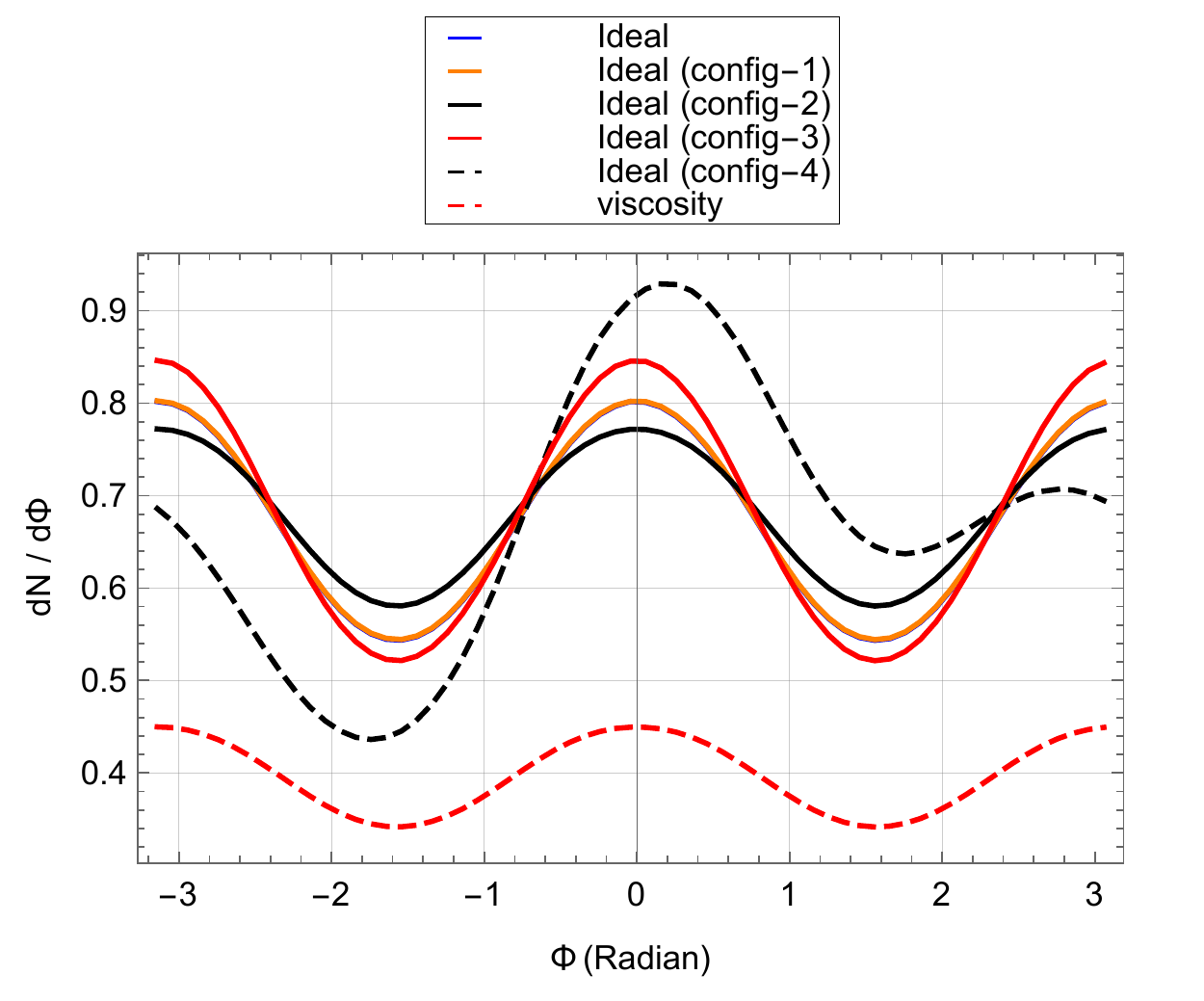} \\
		\caption{(Color online) Top panel: $dN/d\phi$ as a function of $\phi$ for $\pi^{+}$ for various field configurations. Bottom panel: same as top panel but for proton. } {\label{dndphiplotpionproton}}
	\end{figure}
	
	Before discussing the differential $v_2$ for different cases, it is worthwhile to explore the dependence of $p_T$ integrated (0-3 GeV) $dN/d\phi$ on different field configurations and viscosity. In the top panel of Fig.(\ref{dndphiplotpionproton}) we show the $dN/d\phi$ of $\pi^{+}$ as a function of $\phi$ for various cases. As expected we have pure cosine like dependence of $dN/d\phi$ for ideal case (shown by the solid blue line),
the viscosity reduces the multiplicity as well as the amplitude (shown by the red dashed line).  {\bf{config-1}} the isotropic field configuration (orange solid line) almost coincides with the ideal result, but we see noticeable change in the amplitude of $dN/d\phi$ for  {\bf{config-2}}, {\bf{config-3}} as expected. Things become more interesting for  {\bf{config-4}} (black dashed line), here we generate finite directed flow like behaviour,
this is understood from the fact that a unidirectional force shifts the centre of mass of the distribution. Similar behaviour was observed for protons
also (shown in the bottom panel of Fig.(\ref{dndphiplotpionproton}). 

A more concrete way to study the angular dependence of $dN/d\phi$ for {\bf{config-4}} can be achieved by using a finite Fourier series decomposition of $dN/d\phi$ as shown in Eq.\eqref{fitfunction}.
\begin{equation} {\label{fitfunction}}
		f(\Phi)= N \left[ 1+ 2 \sum_{n=1}^{3} v_n cos (n\Phi) +2 \sum_{n=1}^{3} w_n sin (n\Phi) \right].
\end{equation}
A non-linear least squares fit with $v_n, w_n$ and $N$ as a free-parameters we obtain the best fit for $dN/d\phi$ with the values of these parameters given in Table~\eqref{table:paramset2}. Here we note that the directional force $v_1$ is larger than $v_2$ 
for $\pi^{+}$, and they are similar in magnitude for protons. Moreover, we notice that unlike other cases the azimuthal distribution breaks reflection symmetry with respect to the $Y$ axis which gives rise to non-zero $w_n$ shown in Table~\eqref{table:paramset2}. We also observe a mass
dependence of the directional flow as $\pi^{+}$ has a larger $v_1$ compared to the protons. 

\begin{table}[h] 
		\begin{tabular}{|p{0.5cm}|p{3.5cm}|p{3.5cm}|}
			\hline
			&    $\pi^{+}$& $ {\rm{p}}$   \\ 
			\hline
			$N$   & 11.907 $\pm$  2.7e-05  & 0.672 $\pm$ 7.1e-08 \\
			\hline
			$v_1$   &  0.218 $\pm$ 1.0e-07  &  0.081 $\pm$ 7.9e-08 \\
			\hline
			$v_2$    & 0.064 $\pm$ 9.8e-08 & 0.096 $\pm$ 7.9e-08 \\
			\hline
			$ v_3 $   &  0.017 $\pm$ 9.7e-08 & 0.005 $\pm$ 7.8e-08 \\
			\hline
			$ w_1 $  &  0.213 $\pm$ 1.0e-07 & 0.081 $\pm$ 7.9e-08 \\
			\hline
			$ w_2 $  & 0.000  $\pm$ 9.8e-08  & 0.000  $\pm$ 7.8e-08 \\
			\hline
			$ w_3 $   & 0.017 $\pm$ 9.8e-08 & 0.005 $\pm$ 7.8e-08\\
			\hline
		\end{tabular}
		\\
		\caption{Fit parameters for $\pi^+$ and proton for {\bf{config-4}} from Eq.\eqref{fitfunction}. }
		\label{table:paramset2}
	\end{table}

\begin{figure}[h!]
		\includegraphics[width=.45\textwidth]{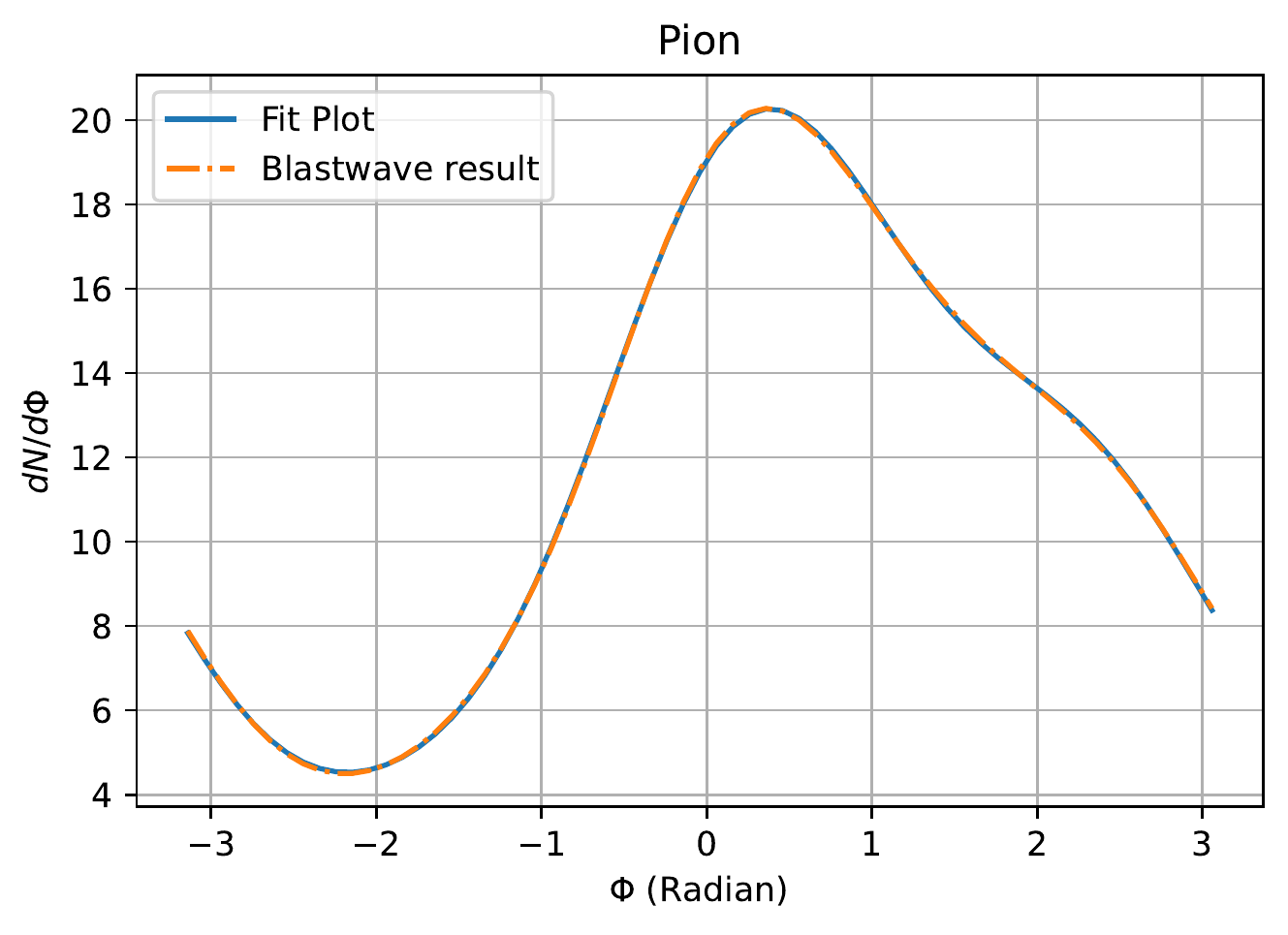}
		\\ 
		\includegraphics[width=.45\textwidth]{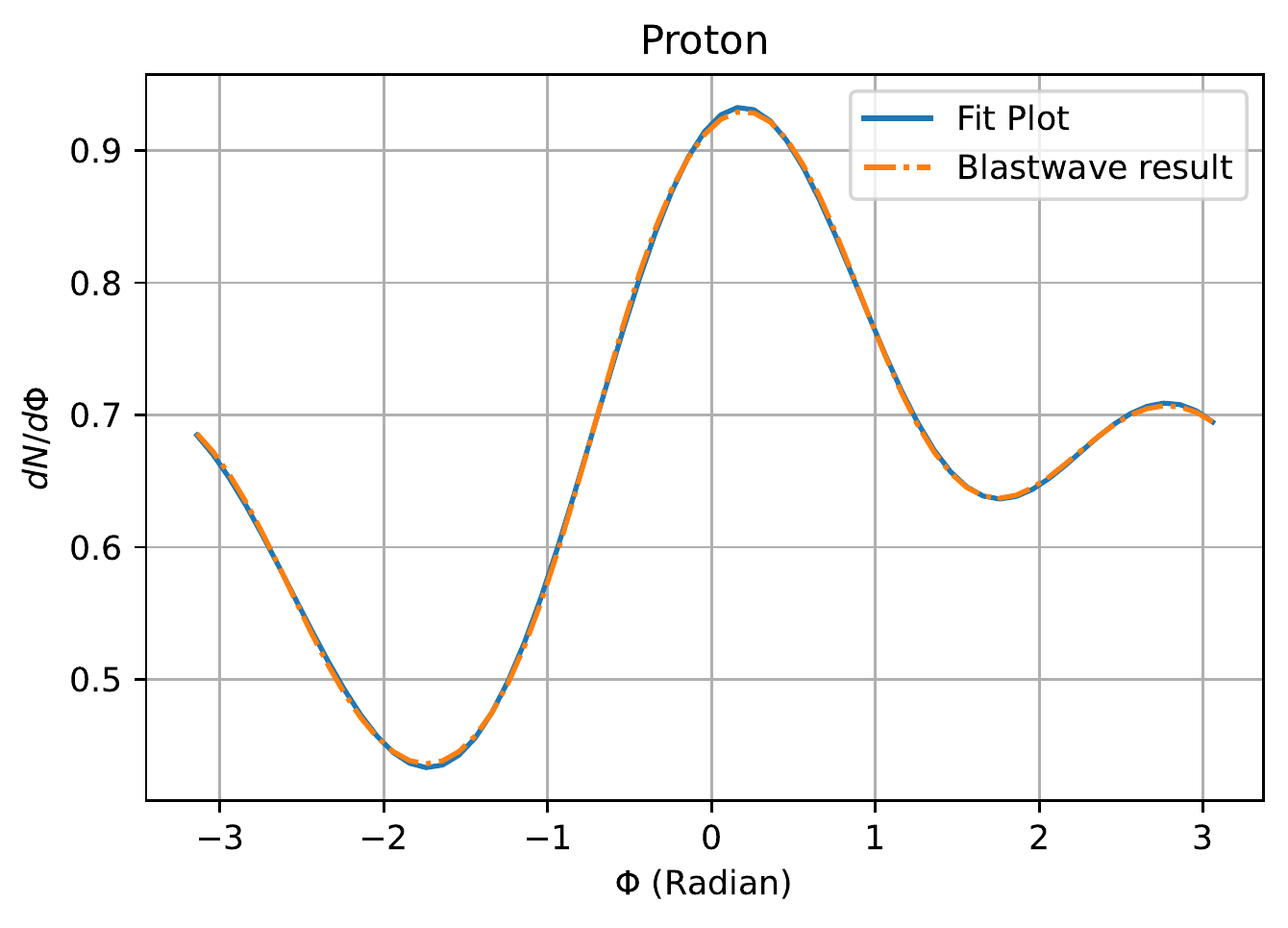}
		\\[\smallskipamount] \hfill
		\caption{(Color online) Blast wave results and the fitted curves using Eq.\eqref{fitfunction} for {\bf{config-4}}  for $\pi^{+}$ (top panel) and proton (bottom panel). }\label{fig:dndphifit}
	\end{figure}

To have a visual understanding of the goodness of fit, we show a comparison of the fitted values (using Eq.\eqref{fitfunction}) (solid blue line) and the $dN/d\phi$ from the blast wave model (dotted-dash orange line) for $\pi^{+}$ (top panel) and 
proton (bottom panel) in Fig.\eqref{fig:dndphifit}. 
	
\begin{figure}[h!]
		\includegraphics[width=0.5\textwidth]{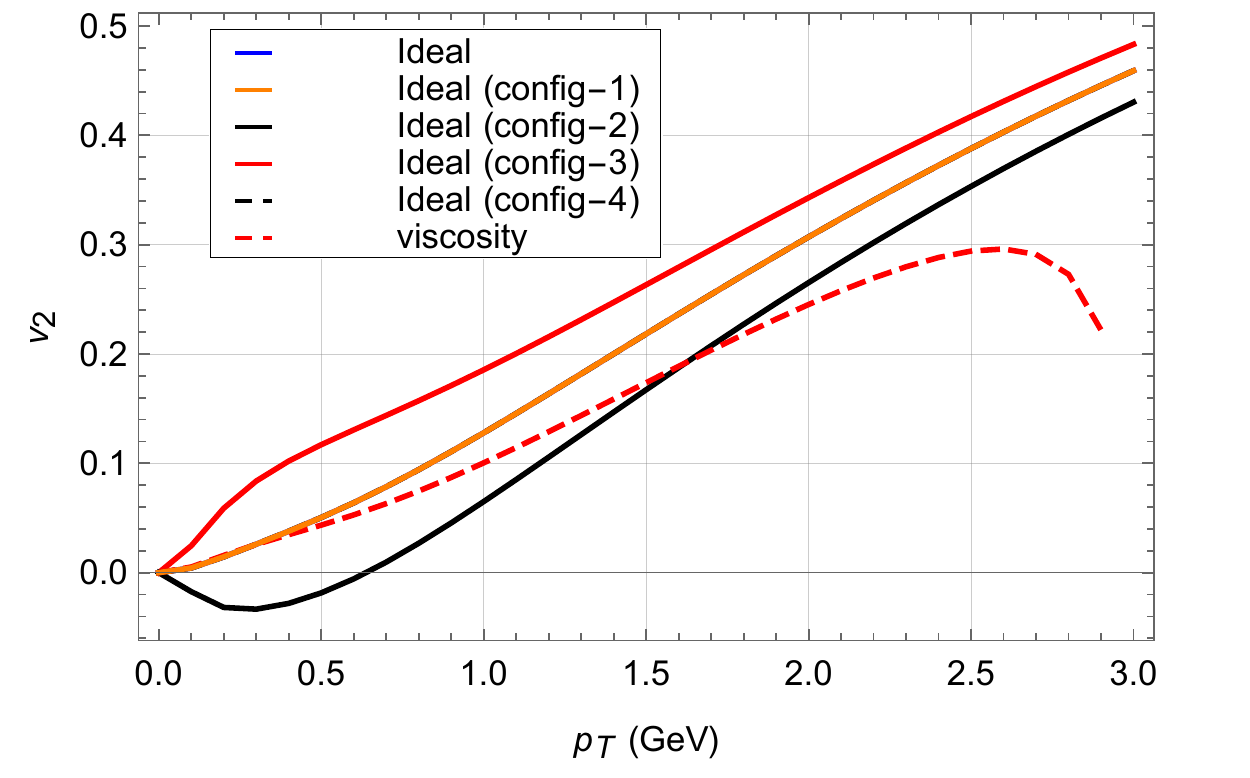} \\ 
		\includegraphics[width=0.5\textwidth]{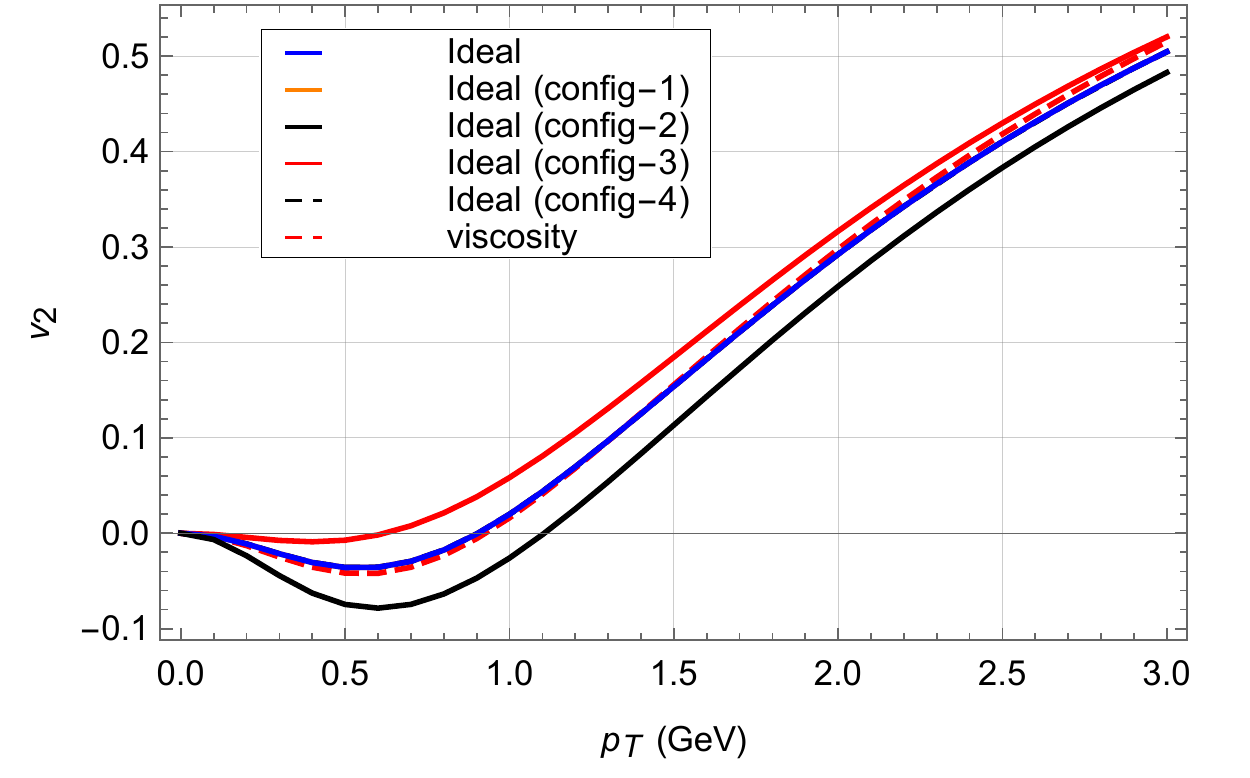}\\ \hfill
		\caption{(Color online) $v_2$ vs $p_{T}$ for $\pi^{+}$ (top panel) and proton (bottom panel) for the transverse electric field configurations shown in Fig.\eqref{electricfieldconfig}.} {\label{pionprotontransverse}}
	\end{figure}

More familiar and useful observables in experiments are centrality and $p_T$ dependent flow harmonics. 
 In Fig.(\ref{pionprotontransverse})  we show the dependence of the second-order flow harmonics $v_2$ (a.k.a elliptic flow) for different configurations. Here we see that there is almost no deviation from the ideal case for the isotropic ({\bf{config-1}}) and directed field case ({\bf{config-4}}). However, the situation is different for  the other two cases, we can clearly see an increase in $v_2$ for {\bf{config-3}} and a suppression for {\bf{config-2}}. We also note that viscosity supresses the elliptic (red dashed line) flow for $\pi^{+}$ and elavates for proton.  

\begin{figure}[h!]
		\includegraphics[width=.5\textwidth,height=0.3\textwidth]{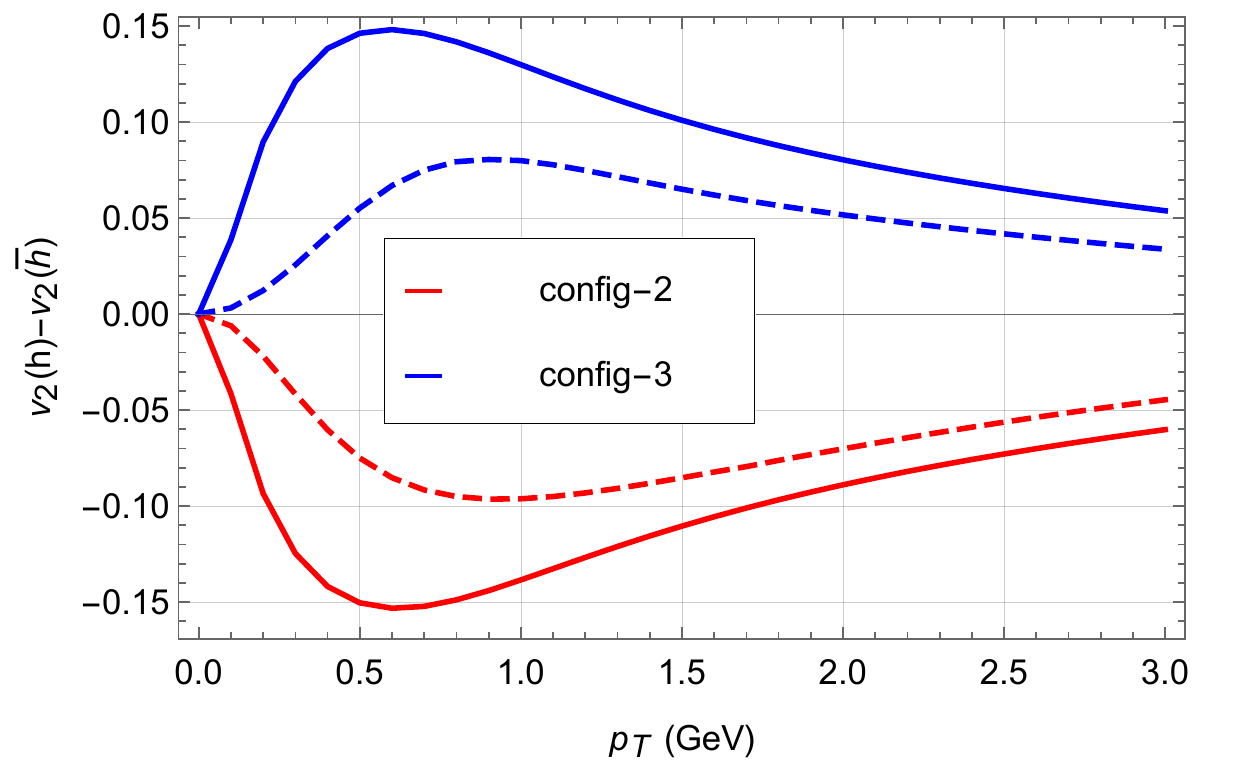}
		\\ \hfill
		\caption{(Color online)  $v_{2}$(${\rm{h}}$) -$v_{2}$(${\rm{\bar{h}}}$) as a function of $p_{T}$ for  {\bf{config-2}} (red) and  {\bf{config-3}} (blue) line corresponds to $\pi$ (Solid line) and $\rm{p}$ (Dashed line). }\label{v2-v2bar}
	\end{figure}

The effect of electric fields become more interesting when we examine the difference in $v_2$ for particles ($\pi^{+},\rm{p}$) and antiparticles 
($\pi^{-},\bar{\rm{p}}$). This difference $\Delta v_2 = v_2(\rm{h})-v_2(\bar{h})$ is shown in Fig.\eqref{v2-v2bar} as a function of $p_T$. We observe 
a non-monotonic variation in $\Delta v_2$ as a function of $p_T$ for both pions and protons. Interestingly, a similar observation was made in~\cite{Inghirami:2019mkc}. 

\begin{figure}[h!]
		\includegraphics[width=.5\textwidth,height=0.17\textwidth]{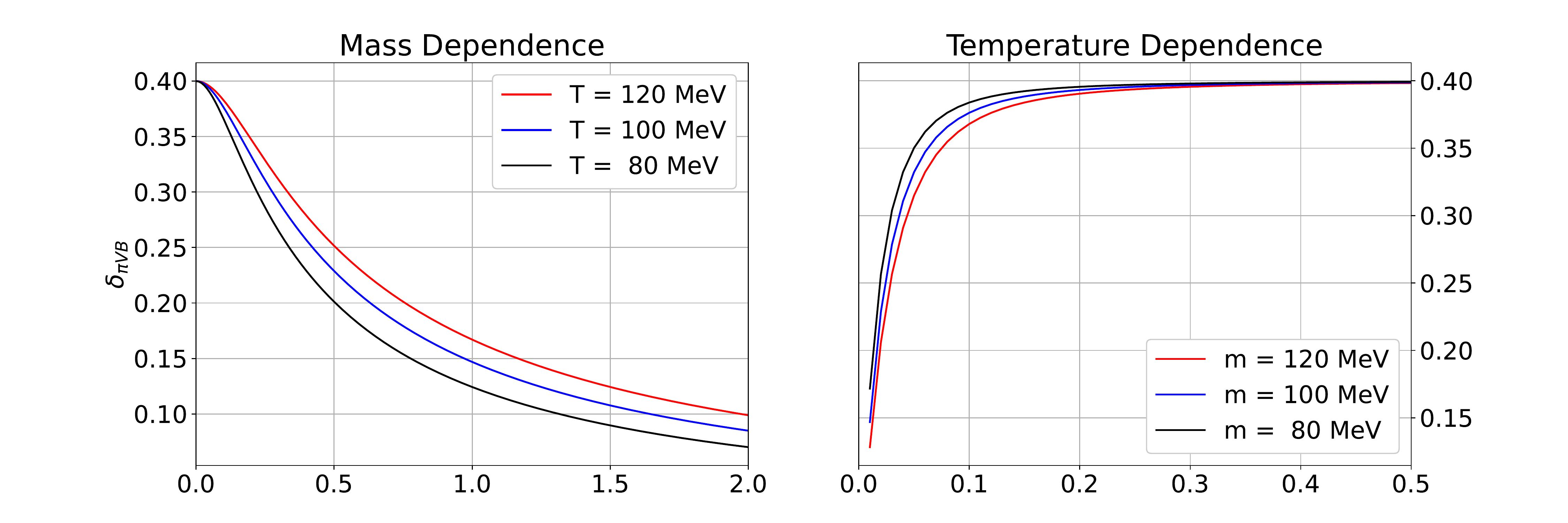}
		\\[\smallskipamount] \hfill
		\includegraphics[width=.5\textwidth,height=0.15\textwidth]{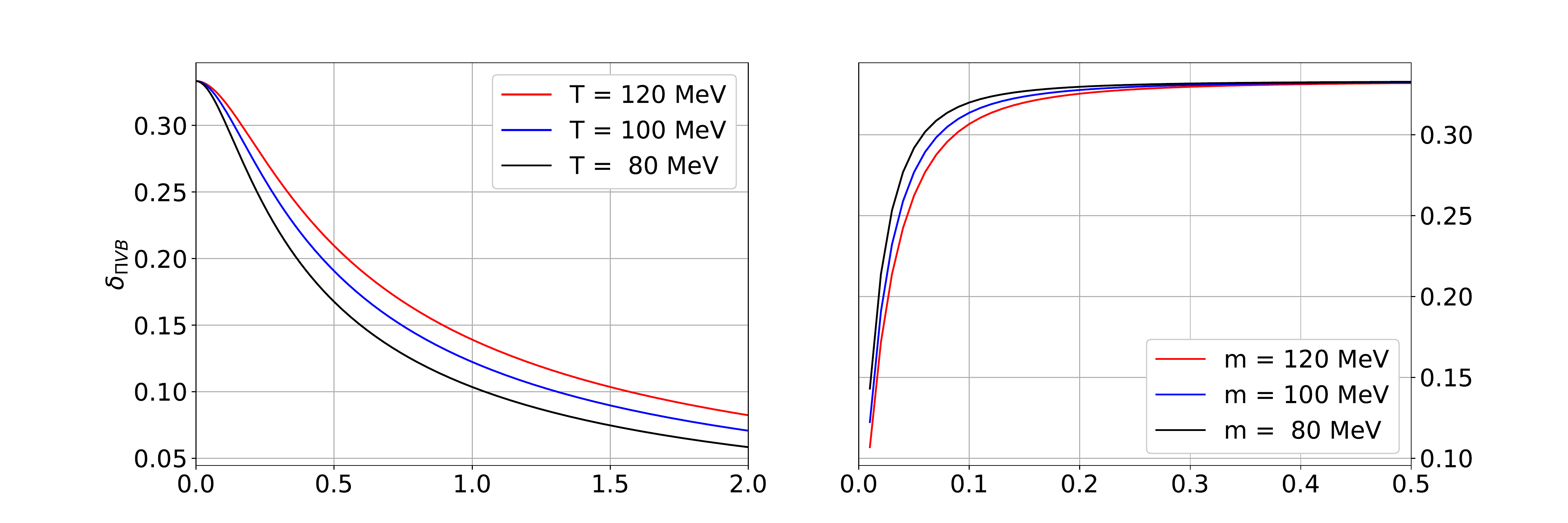}
		\\[\smallskipamount] \hfill
		\includegraphics[width=.5\textwidth,height=0.15\textwidth]{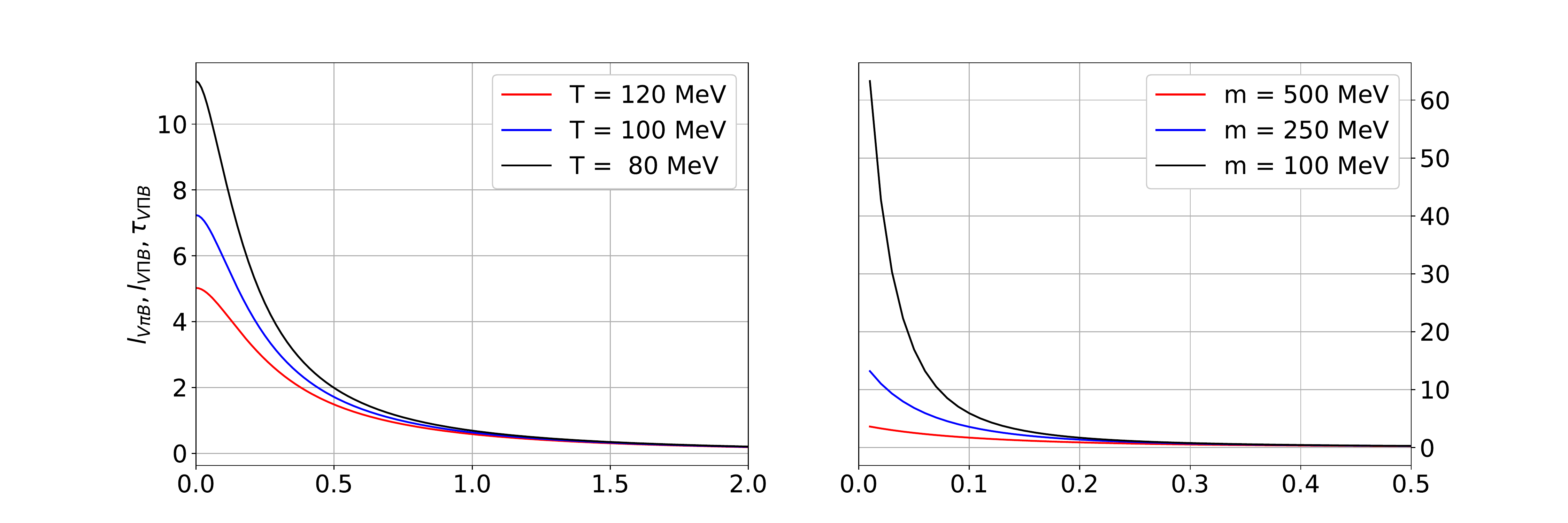}
		\\[\smallskipamount]
		\includegraphics[width=.5\textwidth,height=0.15\textwidth]{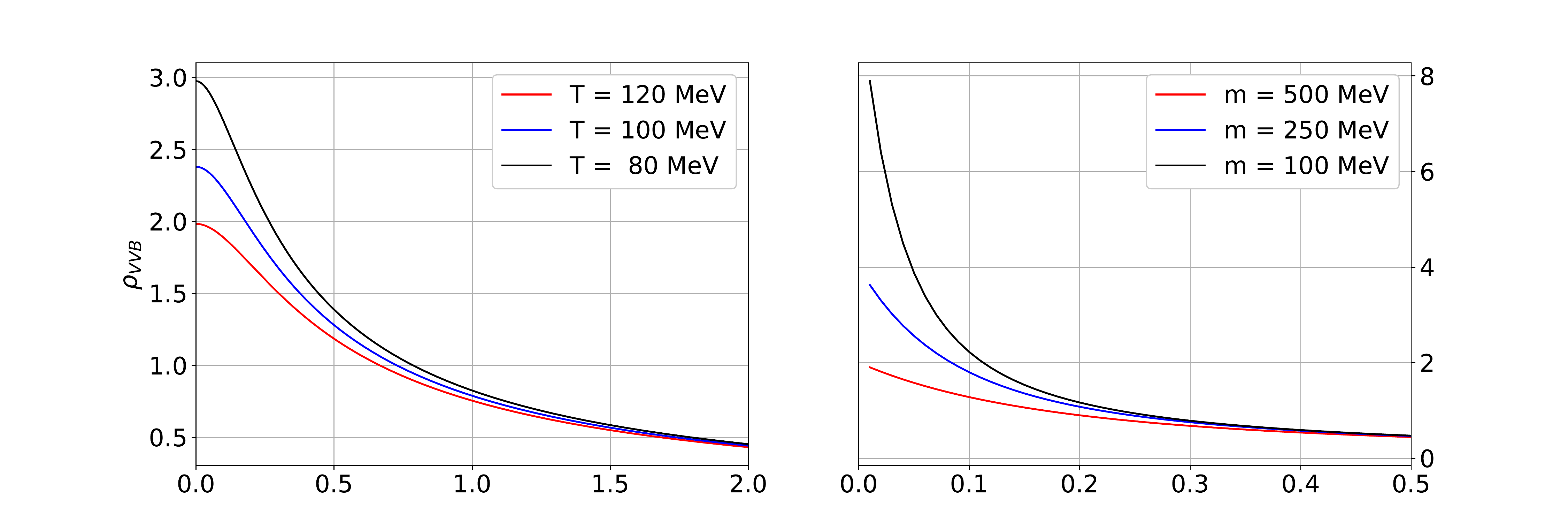}
		\\[\smallskipamount]
		\includegraphics[width=.5\textwidth,height=0.15\textwidth]{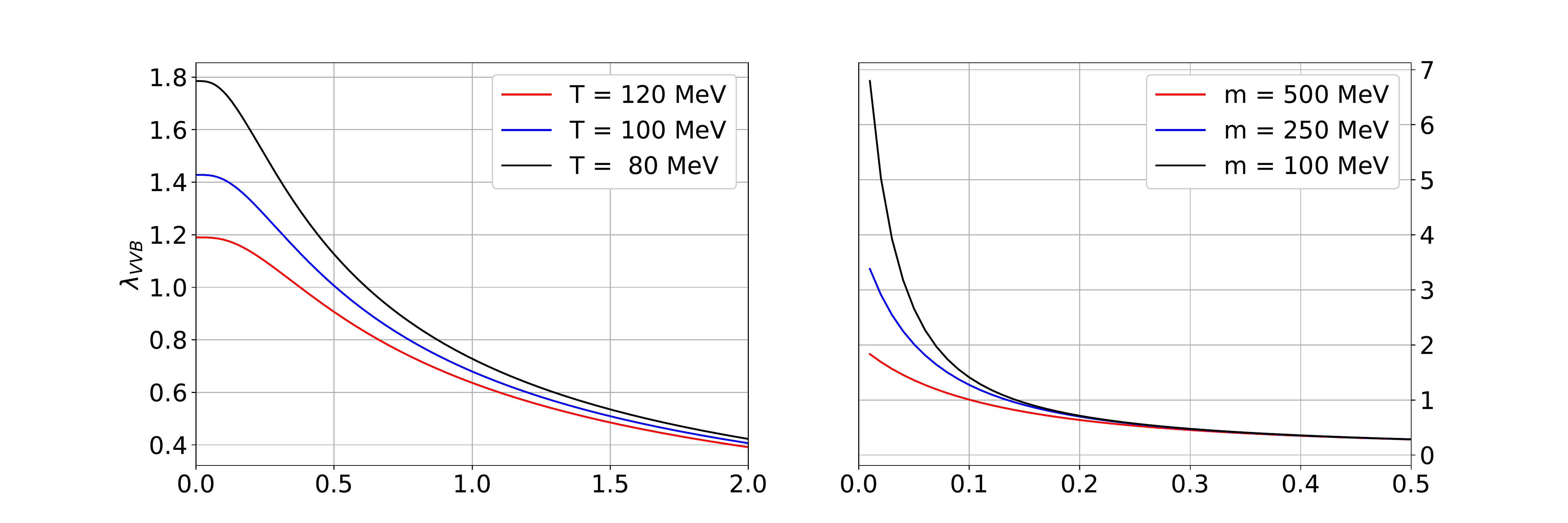}
		\\[\smallskipamount]
		\includegraphics[width=.5\textwidth,height=0.15\textwidth]{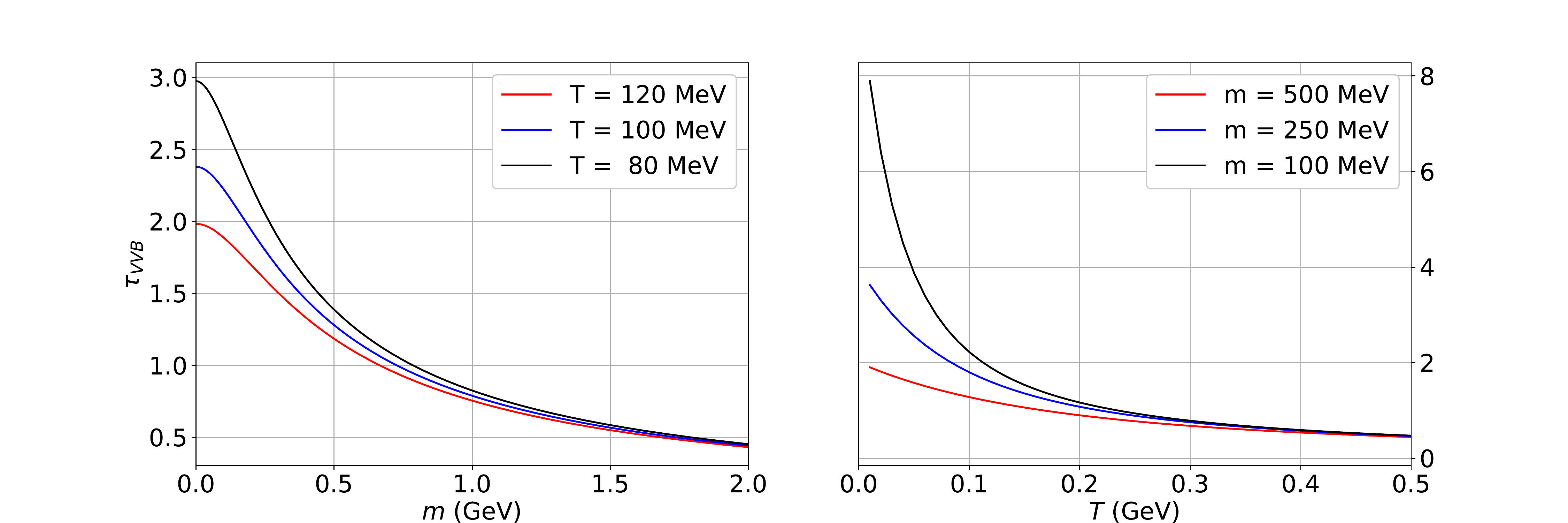}
		\\[\smallskipamount]
		\caption{(Color online)Mass and temperature variation of the transport coefficients arising due to external magnetic field.}\label{fig:TCinMagField}
	\end{figure}

Throughout this study we only consider the effect of electric fields and the first order correction in the $\delta f$. As mentioned earlier, previous studies showed that there are new transport coefficients at higher order corrections to $f$ which may alter the results obtained here.
But these are beyond the scope of the present exploratory study. In Fig.\eqref{fig:TCinMagField} we show the temperature and mass dependence of 
some of the transport coefficients arising due to the magnetic fields. This may give some hints of the relative contribution of various 
transport coefficients in the bulk observables while used in the Cooper-Frye prescription.

%

	\section{Summary and Conclusion}{\label{conclusion}}
	 In this work, we have studied the effect of electric fields on the bulk observables in heavy-ion collisions such as $p_{T}$ spectra, directed and elliptic flow of charged pions and protons. We use the blast-wave model and different configurations of electric fields in the transverse plane to carry out this exploratory study. The $p_{T}$ spectra of hadrons in the blast-wave model are obtained using Cooper-Frye prescription, where we incorporate non-equilibrium correction $\delta f$ due to the viscosity and the electric fields. Since the blast-wave model does not include space-time evolution, the fluid velocity and the electric fields are parameterized on the freezeout hypersurface to 
calculate experimental observables. 	For our case, fluid velocity fields are modulated so that it dominantly generates the elliptic flow. 
We use four different configurations of transverse electric fields (i) isotropic fields, (ii) prolate-like fields, (iii) oblate-like fields, and (iv) directed fields.
The typical maximum value of the electric field for all these configurations is $\sim m_{\pi}^{2}$.
We find that flow harmonics for isotropic fields remain unchanged for both pions and protons. Both prolate and oblate-like field configuration
alters the flow harmonics, and the directed field gives rise to large directed flow $v_1$ for both pions and protons. We also observe a mass
dependence of $v_1$ generated due to the electric fields. 
We also discuss the temperature and mass dependence of some of the new transport coefficients that appear in the second-order correction 
in the distribution function due to the EM field. They can further contribute to the results we obtained here. This we leave for 
a possible future study.
	
	\begin{acknowledgments}
		AP acknowledges the CSIR-HRDG financial support. RG  and VR  acknowledge support from the DAE, Govt. of India.
	\end{acknowledgments}

	\appendix
	\section{Appendix}\label{app:definition}
	In this work we have used Milne coordinate system ($\tau,\eta,r,\phi$), where $\tau=\sqrt{t^2-z^2}$, $r=\sqrt{x^2+y^2}$, $\eta=\tanh^{-1}(z/t)$, and $\phi=\tan^{-1}(y/x)$, due to the co-ordinate transformation various equations changed forms compared to the Cartesian coordinates.
	Here we give the details about the Jacobian and Christoffel symbols used in this study due to the above coordinate transformation:
	the Jacobian for volume element is $\sqrt{-g}=\tau r$ and the non-vanishing Christoffel symbols are
	$\Gamma^{\tau}_{\eta\eta}=\tau$ ,$\Gamma^{\eta}_{\tau\eta}=\frac{1}{\tau}$, $\Gamma^{r}_{\phi\phi}=-r$ ,$\Gamma^{\phi}_{r\phi}=\frac{1}{r}$. The space-like projection is defined as 
	$\Delta^{\mu\nu}=g^{\mu\nu}-u^{\mu}u^{\nu}$ ; for Milne coordinate system, different components of $\Delta^{\mu\nu}$ are
	$\Delta^{\tau\tau}=-\left(u^r\right)^2$ ,
	$\Delta^{\eta\eta}=-\frac{1}{\tau^2}$ ,
	$\Delta^{rr}=-1-\left(u^r\right)^2$ ,
	$\Delta^{\phi\phi}=-\frac{1}{r^2}$ ,
	$\Delta^{\tau\eta}=\Delta^{\tau\phi}=\Delta^{\eta r}=\Delta^{\eta \phi}=\Delta^{r\phi}=0$ , $\Delta^{\tau r}=-\sqrt{1+(u^r)^2}u^{r}$. The expansion scalar is given by 
	$ \theta =D_{\mu} u^{\mu} =  \partial_{\mu} u^{\mu} + \Gamma^{\mu}_{\mu \alpha} u^{\alpha}$ .
	
	We consider real particles and the on-shell condition is given by $g_{\mu\nu}p^{\mu}p^{\nu}=\left(p^{\tau}\right)^2-\tau^2 \left(p^{\eta}\right)^2-\left(p^{r}\right)^2-r^2\left(p^{\phi}\right)^2=m^2$. Following the convention used in heavy-ion collisions we express the components of the 
	four momentum $p^{\mu}$ as $\left(E,p^x,p^y,p^z\right)=\left(m_T \cosh{y},p_T \cos{\varphi},p_{T}\sin{\varphi},m_T \sinh{y}\right)$. Where $p_T=\sqrt{p_x^2+p_y^2}$ , $m_T=\sqrt{m^2+p_T^2}$, and $y= tanh^{-1}\left(p_z/E\right)$. The components of the four-momentum in Milne co-ordinates are
	\begin{eqnarray}\nonumber
		p^{\tau}&=&m_T \cosh{(y-\eta)},\\ \nonumber
		\tau p^{\eta}&=&m_T \sinh{(y- \eta)},\\ \nonumber
		p^r&=&p_{T}\cos{(\varphi-\phi)},\\ \nonumber
		r p^{\phi}&=&p_T \sin{(\varphi-\phi)}.
	\end{eqnarray}
	$\varphi$ is the azimuthal angle of the particle in the momentum space.

In Eq.\eqref{eq:invYield} we have the term $p^{\mu}d\Sigma_{\mu}$ which in our is given by 
	$p^{\mu} d\Sigma_{\mu} = g_{\mu\nu} p^{\mu} d \Sigma^{\nu} = m_T \cosh{ (y -\eta)} \tau d\eta rdr d\phi $. 

\begin{widetext}
		\section{First-order ($\delta f$)  correction to the single-particle distribution } \label{deltaf}	
		
In Eq.\eqref{cooperfrey} we introduced the first-order  correction to the single-particle distribution while calculating invariant yield using 
the Cooper-Frye formula. Here we give the detail expression of $\delta f$ in terms of gradients of fluid variables and fields:
		\begin{eqnarray}\nonumber
			\delta f &=& -\frac{\tau_c}{u.p} \left( p^{\mu} \partial_{\mu} f_0 + q F^{\mu\nu} p_{\nu} \frac{\partial f_0}{\partial p^{\mu}}  \right) \\ \nonumber
			 &=& - \frac{\tau_c}{u.p} \left( -p^{\mu} f_0 \tilde{f}_0 \left[ \beta p^{\alpha} D_{\mu} u_{\alpha} + (u.p) \partial_{\mu} \beta  - \partial_{\mu} \alpha\right]  \right) - \frac{\tau_c}{u.p} \left( - f_0 \tilde{f}_0 q F^{\mu\nu} p_{\nu} \beta \frac{\partial (u.p)}{\partial p^{\mu}}   \right) \\ \nonumber
			 &=& \frac{\tau_c f_0 \tilde{f}_0}{u.p} \left( \beta p^{\mu} p^{\alpha} D_{\mu} u_{\alpha} + (u.p) p^{\mu} \partial_{\mu}\beta  - p^{\mu} \partial_{\mu} \alpha \right)  - \frac{\tau_c f_0 \tilde{f}_0}{u.p} q \beta E^{\nu} p_{\nu} \\ \nonumber
			&=& \frac{\tau_c f_0 \tilde{f}_0}{u.p} \left( \beta p^{\mu} p^{\alpha} \left[ u_{\mu} \dot{u}_{\alpha} + \sigma_{\mu\alpha} + \omega_{\mu\alpha} + \frac{\Delta_{\mu\alpha} \theta}{3} \right] + (u.p) p^{\mu} \partial_{\mu}\beta  - p^{\mu} \partial_{\mu} \alpha \right)  - \frac{\tau_c f_0 \tilde{f}_0}{u.p} q \beta E^{\nu} p_{\nu} \\ \nonumber
			 &=& \frac{\tau_c f_0 \tilde{f}_0}{u.p} \left( \beta p^{\mu} p^{\alpha} \left[  \sigma_{\mu\alpha} + \frac{\Delta_{\mu\alpha} \theta}{3} \right] + (u.p) p^{\mu} \partial_{\mu}\beta  - p^{\mu} \partial_{\mu} \alpha \right)  - \frac{\tau_c f_0 \tilde{f}_0}{u.p} q \beta E^{\nu} p_{\nu} \\ \nonumber
			 &=& \frac{\tau_c f_0 \tilde{f}_0}{u.p} \left( \beta p^{\mu} p^{\alpha} \partial_{\mu} u_{\alpha} + \beta p^{\phi} p^{\phi} r u_{r} -  \beta p^{\eta} p^{\eta} \tau u_{\tau}  + (u.p) p^{\mu} \partial_{\mu}\beta  - p^{\mu} \partial_{\mu} \alpha \right)  - \frac{\tau_c f_0 \tilde{f}_0}{u.p} q \beta E^{\nu} p_{\nu} \\ \nonumber
			 &=& \frac{\tau_c f_0 \tilde{f}_0}{u.p} \left(   \beta p^{\phi} p^{\phi} r u_{r} -   \beta p^{\eta} p^{\eta} \tau u_{\tau}  + (u.p) p^{\mu} \partial_{\mu}\beta  - p^{\mu} \partial_{\mu} \alpha \right) - \frac{\tau_c f_0 \tilde{f}_0}{u.p} q \beta E^{\nu} p_{\nu} \\ \nonumber
			&&   + \frac{\tau_c f_0 \tilde{f}_0}{u.p}\left(\beta p^{\tau} p^{r} \partial_{\tau} u_{r} + \beta p^{r} p^{r} \partial_{r} u_{r} + \beta p^{\phi} p^{r} \partial_{\phi} u_{r}  \right) + \frac{\tau_c f_0 \tilde{f}_0}{u.p} \left(  \beta p^{\tau} p^{\tau} \partial_{\tau} u_{\tau} + \beta p^{r} p^{\tau} \partial_{r} u_{\tau} + \beta p^{\phi} p^{\tau} \partial_{\phi} u_{\tau} \right) \\ \nonumber
			 &=& \frac{\tau_c f_0 \tilde{f}_0 }{u.p} \left( \beta \left(\frac{p_T \sin{(\varphi-\phi)}}{r} \right)^2 r u_r  - \beta   \left( \frac{m_T \sinh{(y- \eta)}}{\tau}  \right)^2 \tau u_{\tau} + (u.p) p^{\mu} \partial_{\mu}\beta  - p^{\mu} \partial_{\mu} \alpha \right) \\ \nonumber
			&& - \frac{\tau_c f_0 \tilde{f}_0}{u.p} q \beta E^{\nu} p_{\nu}  + \frac{\tau_c f_0 \tilde{f}_0}{u.p}\left(\beta m_T cosh(y-\eta) p_{T} cos(\varphi-\phi) \frac{(u^r)^2}{r u^{\tau}} - \beta (p_T cos (\varphi -\phi))^2 \frac{u^r}{r}    \right)\\ \nonumber
			&&  + \frac{\tau_c f_0 \tilde{f}_0}{u.p} \beta \frac{p_T sin(\varphi -\phi)}{r} p_T cos(\varphi-\phi) u_0 \frac{ 2  r }{R} \sum n u_n \sin{n\left[\phi-\psi_n\right]} \\ \nonumber
			&&   + \frac{\tau_c f_0 \tilde{f}_0}{u.p} \left( - \beta (m_T cosh(y-\eta))^2  \frac{(u^r)^3}{r (u^{\tau})^2} + \beta  m_T cosh(y-\eta) p_{T} cos(\varphi-\phi)  \frac{(u^r)^2}{r u^{\tau}}   \right) \\ \nonumber
			&& - \frac{\tau_c f_0 \tilde{f}_0}{u.p} \beta \frac{p_T sin(\varphi-\phi)}{r} m_T cosh(y-\eta)  \frac{u^r}{ u^{\tau}}  u_0 \frac{ 2 r}{R} \sum n u_n \sin{n\left[\phi-\psi_n\right]},
		\end{eqnarray}
		
		  \begin{eqnarray}\nonumber
       \delta f &=& \frac{\tau_c f_0 \tilde{f}_0 }{u.p}\beta \left(  \frac{u_r}{r}  +\frac{ sin(\varphi -\phi)cos(\varphi-\phi)}{r}  u_0 \frac{ 2  r }{R} \sum u_{n} n\sin{n\left[\phi-\psi_n\right]}   \right)p_{T}^{2}  - \frac{\tau_c f_0 \tilde{f}_0}{u.p} q \beta E^{\nu} p_{\nu} \\ \nonumber
       && + \frac{\tau_c f_0 \tilde{f}_0 }{u.p}\beta \left( 2 cosh(y-\eta)cos(\varphi-\phi) \frac{(u^r)^2}{r u^{\tau}}-    cosh(y-\eta) sin(\varphi-\phi) \frac{u^r}{r u^{\tau}}  u_0 \frac{ 2 r}{R} \sum u_{n}n \sin{n\left[\phi-\psi_n\right]}  \right) m_{T}p_{T}   \\ \nonumber
      &&-  \frac{\tau_c f_0 \tilde{f}_0 }{u.p} \beta\left(  \sinh^2{(y- \eta)}\frac{u_{\tau}}{\tau} +  cosh^{2}(y-\eta)  \frac{(u^r)^3}{r (u^{\tau})^2} \right)  m_{T}^{2}+ \frac{\tau_c f_0 \tilde{f}_0 }{u.p}\left((u.p) p^{\mu} \partial_{\mu}\beta  - p^{\mu} \partial_{\mu} \alpha \right),
  \end{eqnarray}
		
where $\beta = \frac{1}{T}$ and $\alpha = \frac{\mu}{T}$ with $\Gamma^{\tau}_{\eta\eta}=\tau $ along with $\Gamma^{r}_{\phi\phi}=-r$. The contribution due to the electric field $E \cdot p$ in the above equation, when expanded, takes the following form
		\begin{eqnarray}\nonumber
			\delta f &=& - \frac{\tau_c f_0 \tilde{f}_0}{u.p} q \beta E\cdot p, \\ \nonumber
			\delta f &=&  - \frac{\tau_c f_0 \tilde{f}_0}{u.p} q \beta \left( E^{\tau} p^{\tau} - \tau^2 E^{\eta} p^{\eta} - E^r p^r - r^2 E^{\phi} p^{\phi}\right).
		\end{eqnarray}

		\subsection{Co-ordinate transformation of Electric four vector} {\label{electricfieldtransform}}
		The electric field components in Milne-coordinates ($E^{\tau}, E^{\eta}, E^{r} , E^{\phi}$) are connected to the Cartesian components  ($E^{t}, E^{x}, E^{y} , E^{z}$) through the following transformation 		

  \begin{gather}\nonumber
		\begin{bmatrix} E^{\tau}  \\   E^{r} \\  E^{\phi} \\  E^{\eta}  \end{bmatrix}
		=
		\begin{bmatrix}
			Cosh[\eta] & 0 & 0 & -Sinh[\eta] \\
			0 & Cos[\phi] & Sin[\phi] & 0 \\
			0 & \frac{-Sin[\phi]}{r} & \frac{Cos[\phi]}{r} & 0 \\
               \frac{-Sinh[\eta]}{\tau} & 0 & 0 & \frac{Cosh[\eta]}{\tau}
		\end{bmatrix}
		\begin{bmatrix} E^{t}  \\ E^{x} \\ E^{y} \\ E^{z} \end{bmatrix}.
	\end{gather}
		We also note $E.u =0$ and this gives rise to :
		
		\begin{eqnarray}\nonumber
			E^t &=& \frac{E^z Sinh \eta}{cosh \eta} + \frac{(cos \phi E^x + sin \phi E^y) u^r}{cosh \eta u^{\tau}}\,, \\ \nonumber
			E^{\eta}&=& \frac{E^z}{\tau Cosh \eta} - \frac{tanh \eta E^r u^r}{\tau u^{\tau}}\,.
		\end{eqnarray}
		
		As mentioned in the main text we use four different configuration of transverse electric fields, they are parameterised as : 
		\begin{eqnarray} \nonumber
			e E_x &=& \frac{ \mathcal{B} Z \alpha_{em} (x - x_0)
				Cosh[\eta - \eta_0]}{((x - x_0)^2 + (y - y_0)^2 + (\tau
				Sinh[\eta - \eta_0])^2)^{3/2}} , \\ \nonumber
			eEy &=& \frac{ \mathcal{A} Z \alpha_{em} (y - y_0)
				Cosh[\eta - \eta_0]}{((x - x_0)^2 + (y - y_0)^2 + (\tau
				Sinh[\eta - \eta_0])^2)^{3/2}}, \\ \nonumber
			eEz &=& 0 ,\\ \nonumber
		\end{eqnarray}
		where $Z$ is the atomic number (for our case we choose Z=82), $\alpha_{em}$=$\frac{1}{137}$, $\mathcal{A}$ and $\mathcal{B}$ are the modulation factors which controls the spatial configuration of the field in the transverse plane, and $\alpha$ is the fine structure constant. We get the first 3 configurations in Fig.\eqref{electricfieldconfig} by choosing (i) $\mathcal{A}$=$\mathcal{B}$=10 for {\bf{config-1}}, (ii) $\mathcal{A}$=20 , $\mathcal{B}$=1 for  {\bf{config-2}} (iii) $\mathcal{A}$=1 , $\mathcal{B}$=20 for {\bf{config-3}},  and $q E_x = qE_y = m_{\pi}^2$ for {\bf{config-4}}.

	\end{widetext}

	\bibliography{ref}

\begin{thebibliography}{38}%
\makeatletter
\providecommand \@ifxundefined [1]{%
 \@ifx{#1\undefined}
}%
\providecommand \@ifnum [1]{%
 \ifnum #1\expandafter \@firstoftwo
 \else \expandafter \@secondoftwo
 \fi
}%
\providecommand \@ifx [1]{%
 \ifx #1\expandafter \@firstoftwo
 \else \expandafter \@secondoftwo
 \fi
}%
\providecommand \natexlab [1]{#1}%
\providecommand \enquote  [1]{``#1''}%
\providecommand \bibnamefont  [1]{#1}%
\providecommand \bibfnamefont [1]{#1}%
\providecommand \citenamefont [1]{#1}%
\providecommand \href@noop [0]{\@secondoftwo}%
\providecommand \href [0]{\begingroup \@sanitize@url \@href}%
\providecommand \@href[1]{\@@startlink{#1}\@@href}%
\providecommand \@@href[1]{\endgroup#1\@@endlink}%
\providecommand \@sanitize@url [0]{\catcode `\\12\catcode `\$12\catcode
  `\&12\catcode `\#12\catcode `\^12\catcode `\_12\catcode `\%12\relax}%
\providecommand \@@startlink[1]{}%
\providecommand \@@endlink[0]{}%
\providecommand \url  [0]{\begingroup\@sanitize@url \@url }%
\providecommand \@url [1]{\endgroup\@href {#1}{\urlprefix }}%
\providecommand \urlprefix  [0]{URL }%
\providecommand \Eprint [0]{\href }%
\providecommand \doibase [0]{https://doi.org/}%
\providecommand \selectlanguage [0]{\@gobble}%
\providecommand \bibinfo  [0]{\@secondoftwo}%
\providecommand \bibfield  [0]{\@secondoftwo}%
\providecommand \translation [1]{[#1]}%
\providecommand \BibitemOpen [0]{}%
\providecommand \bibitemStop [0]{}%
\providecommand \bibitemNoStop [0]{.\EOS\space}%
\providecommand \EOS [0]{\spacefactor3000\relax}%
\providecommand \BibitemShut  [1]{\csname bibitem#1\endcsname}%
\let\auto@bib@innerbib\@empty
\bibitem [{\citenamefont {Wilson}(1974)}]{PhysRevD.10.2445}%
  \BibitemOpen
  \bibfield  {author} {\bibinfo {author} {\bibfnamefont {K.~G.}\ \bibnamefont
  {Wilson}},\ }\bibfield  {title} {\bibinfo {title} {Confinement of quarks},\
  }\href {https://doi.org/10.1103/PhysRevD.10.2445} {\bibfield  {journal}
  {\bibinfo  {journal} {Phys. Rev. D}\ }\textbf {\bibinfo {volume} {10}},\
  \bibinfo {pages} {2445} (\bibinfo {year} {1974})}\BibitemShut {NoStop}%
\bibitem [{\citenamefont {Karsch}\ \emph {et~al.}(2000)\citenamefont {Karsch},
  \citenamefont {Laermann},\ and\ \citenamefont {Peikert}}]{Karsch_2000}%
  \BibitemOpen
  \bibfield  {author} {\bibinfo {author} {\bibfnamefont {F.}~\bibnamefont
  {Karsch}}, \bibinfo {author} {\bibfnamefont {E.}~\bibnamefont {Laermann}},\
  and\ \bibinfo {author} {\bibfnamefont {A.}~\bibnamefont {Peikert}},\
  }\bibfield  {title} {\bibinfo {title} {The pressure in 2, 2+1 and 3 flavour
  {QCD}},\ }\href {https://doi.org/10.1016/s0370-2693(00)00292-6} {\bibfield
  {journal} {\bibinfo  {journal} {Physics Letters B}\ }\textbf {\bibinfo
  {volume} {478}},\ \bibinfo {pages} {447} (\bibinfo {year}
  {2000})}\BibitemShut {NoStop}%
\bibitem [{\citenamefont {Romatschke}\ and\ \citenamefont
  {Romatschke}(2007)}]{PhysRevLett.99.172301}%
  \BibitemOpen
  \bibfield  {author} {\bibinfo {author} {\bibfnamefont {P.}~\bibnamefont
  {Romatschke}}\ and\ \bibinfo {author} {\bibfnamefont {U.}~\bibnamefont
  {Romatschke}},\ }\bibfield  {title} {\bibinfo {title} {Viscosity information
  from relativistic nuclear collisions: How perfect is the fluid observed at
  rhic?},\ }\href {https://doi.org/10.1103/PhysRevLett.99.172301} {\bibfield
  {journal} {\bibinfo  {journal} {Phys. Rev. Lett.}\ }\textbf {\bibinfo
  {volume} {99}},\ \bibinfo {pages} {172301} (\bibinfo {year}
  {2007})}\BibitemShut {NoStop}%
\bibitem [{\citenamefont {Song}\ and\ \citenamefont
  {Heinz}(2008)}]{PhysRevC.77.064901}%
  \BibitemOpen
  \bibfield  {author} {\bibinfo {author} {\bibfnamefont {H.}~\bibnamefont
  {Song}}\ and\ \bibinfo {author} {\bibfnamefont {U.}~\bibnamefont {Heinz}},\
  }\bibfield  {title} {\bibinfo {title} {Causal viscous hydrodynamics in 2 + 1
  dimensions for relativistic heavy-ion collisions},\ }\href
  {https://doi.org/10.1103/PhysRevC.77.064901} {\bibfield  {journal} {\bibinfo
  {journal} {Phys. Rev. C}\ }\textbf {\bibinfo {volume} {77}},\ \bibinfo
  {pages} {064901} (\bibinfo {year} {2008})}\BibitemShut {NoStop}%
\bibitem [{\citenamefont {Kovtun}\ \emph {et~al.}(2005)\citenamefont {Kovtun},
  \citenamefont {Son},\ and\ \citenamefont
  {Starinets}}]{PhysRevLett.94.111601}%
  \BibitemOpen
  \bibfield  {author} {\bibinfo {author} {\bibfnamefont {P.~K.}\ \bibnamefont
  {Kovtun}}, \bibinfo {author} {\bibfnamefont {D.~T.}\ \bibnamefont {Son}},\
  and\ \bibinfo {author} {\bibfnamefont {A.~O.}\ \bibnamefont {Starinets}},\
  }\bibfield  {title} {\bibinfo {title} {Viscosity in strongly interacting
  quantum field theories from black hole physics},\ }\href
  {https://doi.org/10.1103/PhysRevLett.94.111601} {\bibfield  {journal}
  {\bibinfo  {journal} {Phys. Rev. Lett.}\ }\textbf {\bibinfo {volume} {94}},\
  \bibinfo {pages} {111601} (\bibinfo {year} {2005})}\BibitemShut {NoStop}%
\bibitem [{\citenamefont {Kharzeev}\ \emph {et~al.}(2008)\citenamefont
  {Kharzeev}, \citenamefont {McLerran},\ and\ \citenamefont
  {Warringa}}]{Kharzeev:2007jp}%
  \BibitemOpen
  \bibfield  {author} {\bibinfo {author} {\bibfnamefont {D.~E.}\ \bibnamefont
  {Kharzeev}}, \bibinfo {author} {\bibfnamefont {L.~D.}\ \bibnamefont
  {McLerran}},\ and\ \bibinfo {author} {\bibfnamefont {H.~J.}\ \bibnamefont
  {Warringa}},\ }\bibfield  {title} {\bibinfo {title} {{The Effects of
  topological charge change in heavy ion collisions: 'Event by event P and CP
  violation'}},\ }\href {https://doi.org/10.1016/j.nuclphysa.2008.02.298}
  {\bibfield  {journal} {\bibinfo  {journal} {Nucl. Phys. A}\ }\textbf
  {\bibinfo {volume} {803}},\ \bibinfo {pages} {227} (\bibinfo {year}
  {2008})},\ \Eprint {https://arxiv.org/abs/0711.0950} {arXiv:0711.0950
  [hep-ph]} \BibitemShut {NoStop}%
\bibitem [{\citenamefont {Skokov}\ \emph {et~al.}(2009)\citenamefont {Skokov},
  \citenamefont {Illarionov},\ and\ \citenamefont {Toneev}}]{Skokov:2009qp}%
  \BibitemOpen
  \bibfield  {author} {\bibinfo {author} {\bibfnamefont {V.}~\bibnamefont
  {Skokov}}, \bibinfo {author} {\bibfnamefont {A.~Y.}\ \bibnamefont
  {Illarionov}},\ and\ \bibinfo {author} {\bibfnamefont {V.}~\bibnamefont
  {Toneev}},\ }\bibfield  {title} {\bibinfo {title} {{Estimate of the magnetic
  field strength in heavy-ion collisions}},\ }\href
  {https://doi.org/10.1142/S0217751X09047570} {\bibfield  {journal} {\bibinfo
  {journal} {Int. J. Mod. Phys. A}\ }\textbf {\bibinfo {volume} {24}},\
  \bibinfo {pages} {5925} (\bibinfo {year} {2009})},\ \Eprint
  {https://arxiv.org/abs/0907.1396} {arXiv:0907.1396 [nucl-th]} \BibitemShut
  {NoStop}%
\bibitem [{\citenamefont {Tuchin}(2011)}]{PhysRevC.83.039903}%
  \BibitemOpen
  \bibfield  {author} {\bibinfo {author} {\bibfnamefont {K.}~\bibnamefont
  {Tuchin}},\ }\bibfield  {title} {\bibinfo {title} {Erratum: Synchrotron
  radiation by fast fermions in heavy-ion collisions [phys. rev. c 82, 034904
  (2010)]},\ }\href {https://doi.org/10.1103/PhysRevC.83.039903} {\bibfield
  {journal} {\bibinfo  {journal} {Phys. Rev. C}\ }\textbf {\bibinfo {volume}
  {83}},\ \bibinfo {pages} {039903} (\bibinfo {year} {2011})}\BibitemShut
  {NoStop}%
\bibitem [{\citenamefont {Voronyuk}\ \emph {et~al.}(2011)\citenamefont
  {Voronyuk}, \citenamefont {Toneev}, \citenamefont {Cassing}, \citenamefont
  {Bratkovskaya}, \citenamefont {Konchakovski},\ and\ \citenamefont
  {Voloshin}}]{PhysRevC.83.054911}%
  \BibitemOpen
  \bibfield  {author} {\bibinfo {author} {\bibfnamefont {V.}~\bibnamefont
  {Voronyuk}}, \bibinfo {author} {\bibfnamefont {V.~D.}\ \bibnamefont
  {Toneev}}, \bibinfo {author} {\bibfnamefont {W.}~\bibnamefont {Cassing}},
  \bibinfo {author} {\bibfnamefont {E.~L.}\ \bibnamefont {Bratkovskaya}},
  \bibinfo {author} {\bibfnamefont {V.~P.}\ \bibnamefont {Konchakovski}},\ and\
  \bibinfo {author} {\bibfnamefont {S.~A.}\ \bibnamefont {Voloshin}},\
  }\bibfield  {title} {\bibinfo {title} {Electromagnetic field evolution in
  relativistic heavy-ion collisions},\ }\href
  {https://doi.org/10.1103/PhysRevC.83.054911} {\bibfield  {journal} {\bibinfo
  {journal} {Phys. Rev. C}\ }\textbf {\bibinfo {volume} {83}},\ \bibinfo
  {pages} {054911} (\bibinfo {year} {2011})}\BibitemShut {NoStop}%
\bibitem [{\citenamefont {Deng}\ and\ \citenamefont
  {Huang}(2012)}]{PhysRevC.85.044907}%
  \BibitemOpen
  \bibfield  {author} {\bibinfo {author} {\bibfnamefont {W.-T.}\ \bibnamefont
  {Deng}}\ and\ \bibinfo {author} {\bibfnamefont {X.-G.}\ \bibnamefont
  {Huang}},\ }\bibfield  {title} {\bibinfo {title} {Event-by-event generation
  of electromagnetic fields in heavy-ion collisions},\ }\href
  {https://doi.org/10.1103/PhysRevC.85.044907} {\bibfield  {journal} {\bibinfo
  {journal} {Phys. Rev. C}\ }\textbf {\bibinfo {volume} {85}},\ \bibinfo
  {pages} {044907} (\bibinfo {year} {2012})}\BibitemShut {NoStop}%
\bibitem [{\citenamefont {Tuchin}(2013)}]{Tuchin:2013ie}%
  \BibitemOpen
  \bibfield  {author} {\bibinfo {author} {\bibfnamefont {K.}~\bibnamefont
  {Tuchin}},\ }\bibfield  {title} {\bibinfo {title} {{Particle production in
  strong electromagnetic fields in relativistic heavy-ion collisions}},\ }\href
  {https://doi.org/10.1155/2013/490495} {\bibfield  {journal} {\bibinfo
  {journal} {Adv. High Energy Phys.}\ }\textbf {\bibinfo {volume} {2013}},\
  \bibinfo {pages} {490495} (\bibinfo {year} {2013})},\ \Eprint
  {https://arxiv.org/abs/1301.0099} {arXiv:1301.0099 [hep-ph]} \BibitemShut
  {NoStop}%
\bibitem [{\citenamefont {McLerran}\ and\ \citenamefont
  {Skokov}(2014)}]{McLerran:2013hla}%
  \BibitemOpen
  \bibfield  {author} {\bibinfo {author} {\bibfnamefont {L.}~\bibnamefont
  {McLerran}}\ and\ \bibinfo {author} {\bibfnamefont {V.}~\bibnamefont
  {Skokov}},\ }\bibfield  {title} {\bibinfo {title} {{Comments About the
  Electromagnetic Field in Heavy-Ion Collisions}},\ }\href
  {https://doi.org/10.1016/j.nuclphysa.2014.05.008} {\bibfield  {journal}
  {\bibinfo  {journal} {Nucl. Phys. A}\ }\textbf {\bibinfo {volume} {929}},\
  \bibinfo {pages} {184} (\bibinfo {year} {2014})},\ \Eprint
  {https://arxiv.org/abs/1305.0774} {arXiv:1305.0774 [hep-ph]} \BibitemShut
  {NoStop}%
\bibitem [{\citenamefont {Israel}\ and\ \citenamefont
  {Stewart}(1979)}]{ISRAEL1979341}%
  \BibitemOpen
  \bibfield  {author} {\bibinfo {author} {\bibfnamefont {W.}~\bibnamefont
  {Israel}}\ and\ \bibinfo {author} {\bibfnamefont {J.}~\bibnamefont
  {Stewart}},\ }\bibfield  {title} {\bibinfo {title} {Transient relativistic
  thermodynamics and kinetic theory},\ }\href
  {https://doi.org/https://doi.org/10.1016/0003-4916(79)90130-1} {\bibfield
  {journal} {\bibinfo  {journal} {Annals of Physics}\ }\textbf {\bibinfo
  {volume} {118}},\ \bibinfo {pages} {341} (\bibinfo {year}
  {1979})}\BibitemShut {NoStop}%
\bibitem [{\citenamefont {Denicol}\ \emph {et~al.}(2018)\citenamefont
  {Denicol}, \citenamefont {Huang}, \citenamefont {Moln\'ar}, \citenamefont
  {Monteiro}, \citenamefont {Niemi}, \citenamefont {Noronha}, \citenamefont
  {Rischke},\ and\ \citenamefont {Wang}}]{Denicol:2018rbw}%
  \BibitemOpen
  \bibfield  {author} {\bibinfo {author} {\bibfnamefont {G.~S.}\ \bibnamefont
  {Denicol}}, \bibinfo {author} {\bibfnamefont {X.-G.}\ \bibnamefont {Huang}},
  \bibinfo {author} {\bibfnamefont {E.}~\bibnamefont {Moln\'ar}}, \bibinfo
  {author} {\bibfnamefont {G.~M.}\ \bibnamefont {Monteiro}}, \bibinfo {author}
  {\bibfnamefont {H.}~\bibnamefont {Niemi}}, \bibinfo {author} {\bibfnamefont
  {J.}~\bibnamefont {Noronha}}, \bibinfo {author} {\bibfnamefont {D.~H.}\
  \bibnamefont {Rischke}},\ and\ \bibinfo {author} {\bibfnamefont
  {Q.}~\bibnamefont {Wang}},\ }\bibfield  {title} {\bibinfo {title}
  {{Nonresistive dissipative magnetohydrodynamics from the Boltzmann equation
  in the 14-moment approximation}},\ }\href
  {https://doi.org/10.1103/PhysRevD.98.076009} {\bibfield  {journal} {\bibinfo
  {journal} {Phys. Rev. D}\ }\textbf {\bibinfo {volume} {98}},\ \bibinfo
  {pages} {076009} (\bibinfo {year} {2018})},\ \Eprint
  {https://arxiv.org/abs/1804.05210} {arXiv:1804.05210 [nucl-th]} \BibitemShut
  {NoStop}%
\bibitem [{\citenamefont {Denicol}\ \emph {et~al.}(2019)\citenamefont
  {Denicol}, \citenamefont {Moln\'ar}, \citenamefont {Niemi},\ and\
  \citenamefont {Rischke}}]{Denicol:2019iyh}%
  \BibitemOpen
  \bibfield  {author} {\bibinfo {author} {\bibfnamefont {G.~S.}\ \bibnamefont
  {Denicol}}, \bibinfo {author} {\bibfnamefont {E.}~\bibnamefont {Moln\'ar}},
  \bibinfo {author} {\bibfnamefont {H.}~\bibnamefont {Niemi}},\ and\ \bibinfo
  {author} {\bibfnamefont {D.~H.}\ \bibnamefont {Rischke}},\ }\bibfield
  {title} {\bibinfo {title} {{Resistive dissipative magnetohydrodynamics from
  the Boltzmann-Vlasov equation}},\ }\href
  {https://doi.org/10.1103/PhysRevD.99.056017} {\bibfield  {journal} {\bibinfo
  {journal} {Phys. Rev. D}\ }\textbf {\bibinfo {volume} {99}},\ \bibinfo
  {pages} {056017} (\bibinfo {year} {2019})},\ \Eprint
  {https://arxiv.org/abs/1902.01699} {arXiv:1902.01699 [nucl-th]} \BibitemShut
  {NoStop}%
\bibitem [{\citenamefont {Panda}\ \emph
  {et~al.}(2021{\natexlab{a}})\citenamefont {Panda}, \citenamefont {Dash},
  \citenamefont {Biswas},\ and\ \citenamefont {Roy}}]{Panda:2020zhr}%
  \BibitemOpen
  \bibfield  {author} {\bibinfo {author} {\bibfnamefont {A.~K.}\ \bibnamefont
  {Panda}}, \bibinfo {author} {\bibfnamefont {A.}~\bibnamefont {Dash}},
  \bibinfo {author} {\bibfnamefont {R.}~\bibnamefont {Biswas}},\ and\ \bibinfo
  {author} {\bibfnamefont {V.}~\bibnamefont {Roy}},\ }\bibfield  {title}
  {\bibinfo {title} {{Relativistic non-resistive viscous magnetohydrodynamics
  from the kinetic theory: a relaxation time approach}},\ }\href
  {https://doi.org/10.1007/JHEP03(2021)216} {\bibfield  {journal} {\bibinfo
  {journal} {JHEP}\ }\textbf {\bibinfo {volume} {03}},\ \bibinfo {pages}
  {216}},\ \Eprint {https://arxiv.org/abs/2011.01606} {arXiv:2011.01606
  [nucl-th]} \BibitemShut {NoStop}%
\bibitem [{\citenamefont {Panda}\ \emph
  {et~al.}(2021{\natexlab{b}})\citenamefont {Panda}, \citenamefont {Dash},
  \citenamefont {Biswas},\ and\ \citenamefont {Roy}}]{Panda:2021pvq}%
  \BibitemOpen
  \bibfield  {author} {\bibinfo {author} {\bibfnamefont {A.~K.}\ \bibnamefont
  {Panda}}, \bibinfo {author} {\bibfnamefont {A.}~\bibnamefont {Dash}},
  \bibinfo {author} {\bibfnamefont {R.}~\bibnamefont {Biswas}},\ and\ \bibinfo
  {author} {\bibfnamefont {V.}~\bibnamefont {Roy}},\ }\bibfield  {title}
  {\bibinfo {title} {{Relativistic resistive dissipative magnetohydrodynamics
  from the relaxation time approximation}},\ }\href
  {https://doi.org/10.1103/PhysRevD.104.054004} {\bibfield  {journal} {\bibinfo
   {journal} {Phys. Rev. D}\ }\textbf {\bibinfo {volume} {104}},\ \bibinfo
  {pages} {054004} (\bibinfo {year} {2021}{\natexlab{b}})},\ \Eprint
  {https://arxiv.org/abs/2104.12179} {arXiv:2104.12179 [nucl-th]} \BibitemShut
  {NoStop}%
\bibitem [{\citenamefont {Panda}\ and\ \citenamefont
  {Roy}(2022)}]{Panda:2022nsw}%
  \BibitemOpen
  \bibfield  {author} {\bibinfo {author} {\bibfnamefont {A.~K.}\ \bibnamefont
  {Panda}}\ and\ \bibinfo {author} {\bibfnamefont {V.}~\bibnamefont {Roy}},\
  }\href@noop {} {\bibinfo {title} {{Wave Phenomena In General Relativistic
  Magnetohydrodynamics}}} (\bibinfo {year} {2022}),\ \Eprint
  {https://arxiv.org/abs/2205.03107} {arXiv:2205.03107 [gr-qc]} \BibitemShut
  {NoStop}%
\bibitem [{\citenamefont {Most}\ and\ \citenamefont
  {Noronha}(2021)}]{Most:2021rhr}%
  \BibitemOpen
  \bibfield  {author} {\bibinfo {author} {\bibfnamefont {E.~R.}\ \bibnamefont
  {Most}}\ and\ \bibinfo {author} {\bibfnamefont {J.}~\bibnamefont {Noronha}},\
  }\bibfield  {title} {\bibinfo {title} {{Dissipative magnetohydrodynamics for
  nonresistive relativistic plasmas: An implicit second-order flux-conservative
  formulation with stiff relaxation}},\ }\href
  {https://doi.org/10.1103/PhysRevD.104.103028} {\bibfield  {journal} {\bibinfo
   {journal} {Phys. Rev. D}\ }\textbf {\bibinfo {volume} {104}},\ \bibinfo
  {pages} {103028} (\bibinfo {year} {2021})},\ \Eprint
  {https://arxiv.org/abs/2109.02796} {arXiv:2109.02796 [astro-ph.HE]}
  \BibitemShut {NoStop}%
\bibitem [{\citenamefont {G\"ursoy}\ \emph {et~al.}(2014)\citenamefont
  {G\"ursoy}, \citenamefont {Kharzeev},\ and\ \citenamefont
  {Rajagopal}}]{PhysRevC.89.054905}%
  \BibitemOpen
  \bibfield  {author} {\bibinfo {author} {\bibfnamefont {U.}~\bibnamefont
  {G\"ursoy}}, \bibinfo {author} {\bibfnamefont {D.}~\bibnamefont {Kharzeev}},\
  and\ \bibinfo {author} {\bibfnamefont {K.}~\bibnamefont {Rajagopal}},\
  }\bibfield  {title} {\bibinfo {title} {Magnetohydrodynamics, charged
  currents, and directed flow in heavy ion collisions},\ }\href
  {https://doi.org/10.1103/PhysRevC.89.054905} {\bibfield  {journal} {\bibinfo
  {journal} {Phys. Rev. C}\ }\textbf {\bibinfo {volume} {89}},\ \bibinfo
  {pages} {054905} (\bibinfo {year} {2014})}\BibitemShut {NoStop}%
\bibitem [{\citenamefont {G\"ursoy}\ \emph {et~al.}(2018)\citenamefont
  {G\"ursoy}, \citenamefont {Kharzeev}, \citenamefont {Marcus}, \citenamefont
  {Rajagopal},\ and\ \citenamefont {Shen}}]{PhysRevC.98.055201}%
  \BibitemOpen
  \bibfield  {author} {\bibinfo {author} {\bibfnamefont {U.}~\bibnamefont
  {G\"ursoy}}, \bibinfo {author} {\bibfnamefont {D.}~\bibnamefont {Kharzeev}},
  \bibinfo {author} {\bibfnamefont {E.}~\bibnamefont {Marcus}}, \bibinfo
  {author} {\bibfnamefont {K.}~\bibnamefont {Rajagopal}},\ and\ \bibinfo
  {author} {\bibfnamefont {C.}~\bibnamefont {Shen}},\ }\bibfield  {title}
  {\bibinfo {title} {Charge-dependent flow induced by magnetic and electric
  fields in heavy ion collisions},\ }\href
  {https://doi.org/10.1103/PhysRevC.98.055201} {\bibfield  {journal} {\bibinfo
  {journal} {Phys. Rev. C}\ }\textbf {\bibinfo {volume} {98}},\ \bibinfo
  {pages} {055201} (\bibinfo {year} {2018})}\BibitemShut {NoStop}%
\bibitem [{\citenamefont {Nakamura}\ \emph {et~al.}(2022)\citenamefont
  {Nakamura}, \citenamefont {Miyoshi}, \citenamefont {Nonaka},\ and\
  \citenamefont {Takahashi}}]{Nakamura:2022wqr}%
  \BibitemOpen
  \bibfield  {author} {\bibinfo {author} {\bibfnamefont {K.}~\bibnamefont
  {Nakamura}}, \bibinfo {author} {\bibfnamefont {T.}~\bibnamefont {Miyoshi}},
  \bibinfo {author} {\bibfnamefont {C.}~\bibnamefont {Nonaka}},\ and\ \bibinfo
  {author} {\bibfnamefont {H.~R.}\ \bibnamefont {Takahashi}},\ }\href@noop {}
  {\bibinfo {title} {{Relativistic resistive magneto-hydrodynamics code for
  high-energy heavy-ion collisions}}} (\bibinfo {year} {2022}),\ \Eprint
  {https://arxiv.org/abs/2211.02310} {arXiv:2211.02310 [nucl-th]} \BibitemShut
  {NoStop}%
\bibitem [{\citenamefont {Inghirami}\ \emph {et~al.}(2016)\citenamefont
  {Inghirami}, \citenamefont {Del~Zanna}, \citenamefont {Beraudo},
  \citenamefont {Moghaddam}, \citenamefont {Becattini},\ and\ \citenamefont
  {Bleicher}}]{Inghirami:2016iru}%
  \BibitemOpen
  \bibfield  {author} {\bibinfo {author} {\bibfnamefont {G.}~\bibnamefont
  {Inghirami}}, \bibinfo {author} {\bibfnamefont {L.}~\bibnamefont
  {Del~Zanna}}, \bibinfo {author} {\bibfnamefont {A.}~\bibnamefont {Beraudo}},
  \bibinfo {author} {\bibfnamefont {M.~H.}\ \bibnamefont {Moghaddam}}, \bibinfo
  {author} {\bibfnamefont {F.}~\bibnamefont {Becattini}},\ and\ \bibinfo
  {author} {\bibfnamefont {M.}~\bibnamefont {Bleicher}},\ }\bibfield  {title}
  {\bibinfo {title} {{Numerical magneto-hydrodynamics for relativistic nuclear
  collisions}},\ }\href {https://doi.org/10.1140/epjc/s10052-016-4516-8}
  {\bibfield  {journal} {\bibinfo  {journal} {Eur. Phys. J. C}\ }\textbf
  {\bibinfo {volume} {76}},\ \bibinfo {pages} {659} (\bibinfo {year} {2016})},\
  \Eprint {https://arxiv.org/abs/1609.03042} {arXiv:1609.03042 [hep-ph]}
  \BibitemShut {NoStop}%
\bibitem [{\citenamefont {Schnedermann}\ \emph {et~al.}(1993)\citenamefont
  {Schnedermann}, \citenamefont {Sollfrank},\ and\ \citenamefont
  {Heinz}}]{Schnedermann:1993ws}%
  \BibitemOpen
  \bibfield  {author} {\bibinfo {author} {\bibfnamefont {E.}~\bibnamefont
  {Schnedermann}}, \bibinfo {author} {\bibfnamefont {J.}~\bibnamefont
  {Sollfrank}},\ and\ \bibinfo {author} {\bibfnamefont {U.~W.}\ \bibnamefont
  {Heinz}},\ }\bibfield  {title} {\bibinfo {title} {{Thermal phenomenology of
  hadrons from 200-A/GeV S+S collisions}},\ }\href
  {https://doi.org/10.1103/PhysRevC.48.2462} {\bibfield  {journal} {\bibinfo
  {journal} {Phys. Rev. C}\ }\textbf {\bibinfo {volume} {48}},\ \bibinfo
  {pages} {2462} (\bibinfo {year} {1993})},\ \Eprint
  {https://arxiv.org/abs/nucl-th/9307020} {arXiv:nucl-th/9307020} \BibitemShut
  {NoStop}%
\bibitem [{\citenamefont {Fukushima}\ \emph {et~al.}(2008)\citenamefont
  {Fukushima}, \citenamefont {Kharzeev},\ and\ \citenamefont
  {Warringa}}]{PhysRevD.78.074033}%
  \BibitemOpen
  \bibfield  {author} {\bibinfo {author} {\bibfnamefont {K.}~\bibnamefont
  {Fukushima}}, \bibinfo {author} {\bibfnamefont {D.~E.}\ \bibnamefont
  {Kharzeev}},\ and\ \bibinfo {author} {\bibfnamefont {H.~J.}\ \bibnamefont
  {Warringa}},\ }\bibfield  {title} {\bibinfo {title} {Chiral magnetic
  effect},\ }\href {https://doi.org/10.1103/PhysRevD.78.074033} {\bibfield
  {journal} {\bibinfo  {journal} {Phys. Rev. D}\ }\textbf {\bibinfo {volume}
  {78}},\ \bibinfo {pages} {074033} (\bibinfo {year} {2008})}\BibitemShut
  {NoStop}%
\bibitem [{\citenamefont {Burnier}\ \emph {et~al.}(2011)\citenamefont
  {Burnier}, \citenamefont {Kharzeev}, \citenamefont {Liao},\ and\
  \citenamefont {Yee}}]{PhysRevLett.107.052303}%
  \BibitemOpen
  \bibfield  {author} {\bibinfo {author} {\bibfnamefont {Y.}~\bibnamefont
  {Burnier}}, \bibinfo {author} {\bibfnamefont {D.~E.}\ \bibnamefont
  {Kharzeev}}, \bibinfo {author} {\bibfnamefont {J.}~\bibnamefont {Liao}},\
  and\ \bibinfo {author} {\bibfnamefont {H.-U.}\ \bibnamefont {Yee}},\
  }\bibfield  {title} {\bibinfo {title} {Chiral magnetic wave at finite baryon
  density and the electric quadrupole moment of the quark-gluon plasma},\
  }\href {https://doi.org/10.1103/PhysRevLett.107.052303} {\bibfield  {journal}
  {\bibinfo  {journal} {Phys. Rev. Lett.}\ }\textbf {\bibinfo {volume} {107}},\
  \bibinfo {pages} {052303} (\bibinfo {year} {2011})}\BibitemShut {NoStop}%
\bibitem [{\citenamefont {Kharzeev}\ and\ \citenamefont
  {Yee}(2011)}]{PhysRevD.83.085007}%
  \BibitemOpen
  \bibfield  {author} {\bibinfo {author} {\bibfnamefont {D.~E.}\ \bibnamefont
  {Kharzeev}}\ and\ \bibinfo {author} {\bibfnamefont {H.-U.}\ \bibnamefont
  {Yee}},\ }\bibfield  {title} {\bibinfo {title} {Chiral magnetic wave},\
  }\href {https://doi.org/10.1103/PhysRevD.83.085007} {\bibfield  {journal}
  {\bibinfo  {journal} {Phys. Rev. D}\ }\textbf {\bibinfo {volume} {83}},\
  \bibinfo {pages} {085007} (\bibinfo {year} {2011})}\BibitemShut {NoStop}%
\bibitem [{\citenamefont {Zhang}\ \emph {et~al.}(2022)\citenamefont {Zhang},
  \citenamefont {Sheng}, \citenamefont {Pu}, \citenamefont {Chen},
  \citenamefont {Peng}, \citenamefont {Wang},\ and\ \citenamefont
  {Wang}}]{Zhang:2022lje}%
  \BibitemOpen
  \bibfield  {author} {\bibinfo {author} {\bibfnamefont {J.-J.}\ \bibnamefont
  {Zhang}}, \bibinfo {author} {\bibfnamefont {X.-L.}\ \bibnamefont {Sheng}},
  \bibinfo {author} {\bibfnamefont {S.}~\bibnamefont {Pu}}, \bibinfo {author}
  {\bibfnamefont {J.-N.}\ \bibnamefont {Chen}}, \bibinfo {author}
  {\bibfnamefont {G.-L.}\ \bibnamefont {Peng}}, \bibinfo {author}
  {\bibfnamefont {J.-G.}\ \bibnamefont {Wang}},\ and\ \bibinfo {author}
  {\bibfnamefont {Q.}~\bibnamefont {Wang}},\ }\href
  {https://doi.org/10.1103/PhysRevResearch.4.033138} {\bibinfo {title}
  {{Charge-dependent directed flows in heavy-ion collisions by
  Boltzmann-Maxwell equations}}} (\bibinfo {year} {2022}),\ \Eprint
  {https://arxiv.org/abs/2201.06171} {arXiv:2201.06171 [hep-ph]} \BibitemShut
  {NoStop}%
\bibitem [{\citenamefont {Inghirami}\ \emph {et~al.}(2020)\citenamefont
  {Inghirami}, \citenamefont {Mace}, \citenamefont {Hirono}, \citenamefont
  {Del~Zanna}, \citenamefont {Kharzeev},\ and\ \citenamefont
  {Bleicher}}]{Inghirami:2019mkc}%
  \BibitemOpen
  \bibfield  {author} {\bibinfo {author} {\bibfnamefont {G.}~\bibnamefont
  {Inghirami}}, \bibinfo {author} {\bibfnamefont {M.}~\bibnamefont {Mace}},
  \bibinfo {author} {\bibfnamefont {Y.}~\bibnamefont {Hirono}}, \bibinfo
  {author} {\bibfnamefont {L.}~\bibnamefont {Del~Zanna}}, \bibinfo {author}
  {\bibfnamefont {D.~E.}\ \bibnamefont {Kharzeev}},\ and\ \bibinfo {author}
  {\bibfnamefont {M.}~\bibnamefont {Bleicher}},\ }\href
  {https://doi.org/10.1140/epjc/s10052-020-7847-4} {\bibinfo {title} {{Magnetic
  fields in heavy ion collisions: flow and charge transport}}} (\bibinfo {year}
  {2020}),\ \Eprint {https://arxiv.org/abs/1908.07605} {arXiv:1908.07605
  [hep-ph]} \BibitemShut {NoStop}%
\bibitem [{\citenamefont {Roy}\ \emph {et~al.}(2017)\citenamefont {Roy},
  \citenamefont {Pu}, \citenamefont {Rezzolla},\ and\ \citenamefont
  {Rischke}}]{Roy:2017yvg}%
  \BibitemOpen
  \bibfield  {author} {\bibinfo {author} {\bibfnamefont {V.}~\bibnamefont
  {Roy}}, \bibinfo {author} {\bibfnamefont {S.}~\bibnamefont {Pu}}, \bibinfo
  {author} {\bibfnamefont {L.}~\bibnamefont {Rezzolla}},\ and\ \bibinfo
  {author} {\bibfnamefont {D.~H.}\ \bibnamefont {Rischke}},\ }\bibfield
  {title} {\bibinfo {title} {{Effect of intense magnetic fields on reduced-MHD
  evolution in $\sqrt{s_{\rm NN}}$ = 200 GeV Au+Au collisions}},\ }\href
  {https://doi.org/10.1103/PhysRevC.96.054909} {\bibfield  {journal} {\bibinfo
  {journal} {Phys. Rev. C}\ }\textbf {\bibinfo {volume} {96}},\ \bibinfo
  {pages} {054909} (\bibinfo {year} {2017})},\ \Eprint
  {https://arxiv.org/abs/1706.05326} {arXiv:1706.05326 [nucl-th]} \BibitemShut
  {NoStop}%
\bibitem [{\citenamefont {Gezhagn}\ and\ \citenamefont
  {Chaubey}(2022)}]{Gezhagn:2021oav}%
  \BibitemOpen
  \bibfield  {author} {\bibinfo {author} {\bibfnamefont {T.}~\bibnamefont
  {Gezhagn}}\ and\ \bibinfo {author} {\bibfnamefont {A.~K.}\ \bibnamefont
  {Chaubey}},\ }\bibfield  {title} {\bibinfo {title} {{Electromagnetic Field
  Evolution in Relativistic Heavy Ion Collision and Its Effect on Flow of
  Particles}},\ }\href {https://doi.org/10.3389/fphy.2021.791108} {\bibfield
  {journal} {\bibinfo  {journal} {Front. in Phys.}\ }\textbf {\bibinfo {volume}
  {9}},\ \bibinfo {pages} {791108} (\bibinfo {year} {2022})},\ \Eprint
  {https://arxiv.org/abs/2107.01467} {arXiv:2107.01467 [nucl-th]} \BibitemShut
  {NoStop}%
\bibitem [{\citenamefont {Huang}(2016)}]{Huang:2015oca}%
  \BibitemOpen
  \bibfield  {author} {\bibinfo {author} {\bibfnamefont {X.-G.}\ \bibnamefont
  {Huang}},\ }\bibfield  {title} {\bibinfo {title} {{Electromagnetic fields and
  anomalous transports in heavy-ion collisions --- A pedagogical review}},\
  }\href {https://doi.org/10.1088/0034-4885/79/7/076302} {\bibfield  {journal}
  {\bibinfo  {journal} {Rept. Prog. Phys.}\ }\textbf {\bibinfo {volume} {79}},\
  \bibinfo {pages} {076302} (\bibinfo {year} {2016})},\ \Eprint
  {https://arxiv.org/abs/1509.04073} {arXiv:1509.04073 [nucl-th]} \BibitemShut
  {NoStop}%
\bibitem [{\citenamefont {Cooper}\ and\ \citenamefont
  {Frye}(1974)}]{PhysRevD.10.186}%
  \BibitemOpen
  \bibfield  {author} {\bibinfo {author} {\bibfnamefont {F.}~\bibnamefont
  {Cooper}}\ and\ \bibinfo {author} {\bibfnamefont {G.}~\bibnamefont {Frye}},\
  }\bibfield  {title} {\bibinfo {title} {Single-particle distribution in the
  hydrodynamic and statistical thermodynamic models of multiparticle
  production},\ }\href {https://doi.org/10.1103/PhysRevD.10.186} {\bibfield
  {journal} {\bibinfo  {journal} {Phys. Rev. D}\ }\textbf {\bibinfo {volume}
  {10}},\ \bibinfo {pages} {186} (\bibinfo {year} {1974})}\BibitemShut
  {NoStop}%
\bibitem [{\citenamefont {Das}\ \emph {et~al.}(2022)\citenamefont {Das} \emph
  {et~al.}}]{Das:2022lqh}%
  \BibitemOpen
  \bibfield  {author} {\bibinfo {author} {\bibfnamefont {S.~K.}\ \bibnamefont
  {Das}} \emph {et~al.},\ }\bibfield  {title} {\bibinfo {title} {{Dynamics of
  Hot QCD Matter -- Current Status and Developments}}\ }(\bibinfo {year}
  {2022})\ \Eprint {https://arxiv.org/abs/2208.13440} {arXiv:2208.13440
  [nucl-th]} \BibitemShut {NoStop}%
\bibitem [{\citenamefont {Abelev}\ \emph {et~al.}(2013)\citenamefont {Abelev}
  \emph {et~al.}}]{ALICE:2013mez}%
  \BibitemOpen
  \bibfield  {author} {\bibinfo {author} {\bibfnamefont {B.}~\bibnamefont
  {Abelev}} \emph {et~al.} (\bibinfo {collaboration} {ALICE}),\ }\bibfield
  {title} {\bibinfo {title} {{Centrality dependence of $\pi$, K, p production
  in Pb-Pb collisions at $\sqrt{s_{NN}}$ = 2.76 TeV}},\ }\href
  {https://doi.org/10.1103/PhysRevC.88.044910} {\bibfield  {journal} {\bibinfo
  {journal} {Phys. Rev. C}\ }\textbf {\bibinfo {volume} {88}},\ \bibinfo
  {pages} {044910} (\bibinfo {year} {2013})},\ \Eprint
  {https://arxiv.org/abs/1303.0737} {arXiv:1303.0737 [hep-ex]} \BibitemShut
  {NoStop}%
\bibitem [{\citenamefont {Roy}\ and\ \citenamefont
  {Chaudhuri}(2010)}]{Roy:2010qg}%
  \BibitemOpen
  \bibfield  {author} {\bibinfo {author} {\bibfnamefont {V.}~\bibnamefont
  {Roy}}\ and\ \bibinfo {author} {\bibfnamefont {A.~K.}\ \bibnamefont
  {Chaudhuri}},\ }\bibfield  {title} {\bibinfo {title} {{Lattice based equation
  of state and transverse momentum spectra of identified particles in ideal and
  viscous hydrodynamics}},\ }\href@noop {} {\bibfield  {journal} {\bibinfo
  {journal} {DAE Symp. Nucl. Phys.}\ }\textbf {\bibinfo {volume} {55}},\
  \bibinfo {pages} {624} (\bibinfo {year} {2010})},\ \Eprint
  {https://arxiv.org/abs/1003.1195} {arXiv:1003.1195 [nucl-th]} \BibitemShut
  {NoStop}%
\bibitem [{\citenamefont {Noronha-Hostler}\ \emph {et~al.}(2014)\citenamefont
  {Noronha-Hostler}, \citenamefont {Noronha},\ and\ \citenamefont
  {Grassi}}]{PhysRevC.90.034907}%
  \BibitemOpen
  \bibfield  {author} {\bibinfo {author} {\bibfnamefont {J.}~\bibnamefont
  {Noronha-Hostler}}, \bibinfo {author} {\bibfnamefont {J.}~\bibnamefont
  {Noronha}},\ and\ \bibinfo {author} {\bibfnamefont {F.}~\bibnamefont
  {Grassi}},\ }\bibfield  {title} {\bibinfo {title} {Bulk viscosity-driven
  suppression of shear viscosity effects on the flow harmonics at energies
  available at the bnl relativistic heavy ion collider},\ }\href
  {https://doi.org/10.1103/PhysRevC.90.034907} {\bibfield  {journal} {\bibinfo
  {journal} {Phys. Rev. C}\ }\textbf {\bibinfo {volume} {90}},\ \bibinfo
  {pages} {034907} (\bibinfo {year} {2014})}\BibitemShut {NoStop}%
\bibitem [{\citenamefont {Denicol}\ \emph {et~al.}(2010)\citenamefont
  {Denicol}, \citenamefont {Kodama},\ and\ \citenamefont
  {Koide}}]{Denicol:2010tr}%
  \BibitemOpen
  \bibfield  {author} {\bibinfo {author} {\bibfnamefont {G.~S.}\ \bibnamefont
  {Denicol}}, \bibinfo {author} {\bibfnamefont {T.}~\bibnamefont {Kodama}},\
  and\ \bibinfo {author} {\bibfnamefont {T.}~\bibnamefont {Koide}},\ }\bibfield
   {title} {\bibinfo {title} {{The effect of shear and bulk viscosities on
  elliptic flow}},\ }\href {https://doi.org/10.1088/0954-3899/37/9/094040}
  {\bibfield  {journal} {\bibinfo  {journal} {J. Phys. G}\ }\textbf {\bibinfo
  {volume} {37}},\ \bibinfo {pages} {094040} (\bibinfo {year} {2010})},\
  \Eprint {https://arxiv.org/abs/1002.2394} {arXiv:1002.2394 [nucl-th]}
  \BibitemShut {NoStop}%
\end{thebibliography}%

\end{document}